\documentclass[aps,prd,nofootinbib,onecolumn,superscriptaddress,preprintnumbers,letterpaper]{revtex4}

\usepackage{amssymb}
\usepackage{amsmath}
\usepackage{epsfig}
\usepackage{hyperref}
\usepackage{comment}
\usepackage{color}
\usepackage{cleveref}
\usepackage{multirow}
\usepackage{capt-of}
\usepackage{siunitx}
\usepackage{graphicx}
\usepackage{gensymb}
\usepackage[caption=false]{subfig}
\usepackage{slashed}
\usepackage{tikz-feynman}
\usepackage{tabularx}

\makeatletter
\def\simgt{\mathrel{\lower2.5pt\vbox{\lineskip=0pt\baselineskip=0pt
           \hbox{$>$}\hbox{$\sim$}}}}
\def\simlt{\mathrel{\lower2.5pt\vbox{\lineskip=0pt\baselineskip=0pt
           \hbox{$<$}\hbox{$\sim$}}}}
\makeatother
\newcommand{\ie}{{\it i.e.}}

\def\lsim{\mathrel{\raise.3ex\hbox{$<$\kern-.75em\lower1ex\hbox{$\sim$}}}}
\def\gsim{\mathrel{\raise.3ex\hbox{$>$\kern-.75em\lower1ex\hbox{$\sim$}}}}
\begin{document}
\title{TASI Lectures on Indirect Searches For Dark Matter}
\author{Dan Hooper}
\affiliation{Fermi National Accelerator Laboratory, Center for Particle Astrophysics, Batavia, IL 60510 \\ 
University of Chicago, Department of Astronomy and Astrophysics, Chicago, IL 60637 \\
University of Chicago, Kavli Institute for Cosmological Physics, Chicago, IL 60637 \\
orcid.org/0000-0001-8837-4127}
\begin{abstract}

In these lectures, I describe a variety of efforts to identify or constrain the identity of dark matter by detecting the annihilation or decay products of these particles, or their effects. After reviewing the motivation for indirect searches, I discuss what we have learned about dark matter from observations of gamma rays, cosmic rays and neutrinos, as well as the cosmic microwave background. Measurements such as these have been used to significantly constrain a wide range of thermal relic dark matter candidates, in particular those with masses below a few hundred GeV. I also discuss a number of anomalies and excesses that have been interpreted as possible signals of dark matter, including the Galactic Center gamma-ray excess, the cosmic-ray antiproton excess, the cosmic-ray positron excess, and the 3.5 keV line. These lectures were originally presented as part of the 2018 Theoretical Advanced Study Institute (TASI) summer school on ``Theory in an Era of Data''. Although intended for advanced graduate students, these lectures may be useful for a wide range of physicists, astrophysicists and astronomers who wish to get an overview of the current state of indirect searches for dark matter.

\end{abstract}

\preprint{FERMILAB-CONF-18-666-A}

\maketitle

\tableofcontents

\newpage

\section{The Origin of Dark Matter and Motivation for Indirect Searches}
\label{sectionone}

Over the past several decades, weakly interacting massive particles (WIMPs) have generally been considered the leading class of candidates for the dark matter of our universe~\cite{Bertone:2016nfn}. With the goal of identifying the particle nature of this substance, experiments have been designed and carried out to detect the interactions of dark matter particles with atoms (direct detection), to produce particles of dark matter in collider environments, and to detect the products of dark matter annihilations or decays (indirect detection). 

Indirect searches for dark matter include efforts to detect the gamma rays, antiprotons, positrons, neutrinos, and other particles that are produced in the annihilations or decays of this substance. Across a wide range of models, the abundance of dark matter that emerged from the early universe is set by the dark matter's self-annihilation cross section. As I will demonstrate below, a stable particle species with a thermally averaged annihilation cross section of $\langle \sigma v \rangle \simeq 2.2 \times 10^{-26}$ cm$^3/$s is predicted to freeze out of thermal equilibrium with an abundance equal to the measured cosmological density of dark matter~\cite{Steigman:2012nb,Kolb:1990vq,Griest:1990kh}. In many simple models, the dark matter is predicted to annihilate with a similar cross section in the modern universe, providing us with an important benchmark and motivation for indirect searches. Within this context, the current era is an exciting one for indirect detection. In particular, gamma ray and cosmic ray searches for dark matter annihilation products have recently become sensitive to dark matter with this benchmark cross section for masses up to around the weak scale, $\mathcal{O}(10^2$ GeV).

\subsection{The Abundance of a Thermal Relic}

Consider a stable particle, $X$, that can annihilate in pairs. Although I intend to identify this state with the dark matter of our universe, this calculation is quite general and applies to a wide range of stable particle species. The evolution of the number density of this species, $n_X$, is described by the following equation:
\begin{equation}
\frac{dn_X}{dt} + 3 H n_{X} = -\langle \sigma v \rangle [n_X^2-(n^{\rm Eq}_X)^2],
\end{equation}
where $H$ is the rate of Hubble expansion, $\langle \sigma v \rangle$ is the thermally averaged value of the annihilation cross section multiplied by the relative velocity of the two particles, and $n^{\rm Eq}_X$ is the equilibrium number density (\ie~the number density that would be predicted if the $X$ population were in chemical equilibrium with the thermal bath). The Hubble rate is given by
\begin{eqnarray}
H &=&  \bigg(\frac{8 \pi \rho}{3 m^2_{\rm Pl}}\bigg)^{1/2} \nonumber \\
&=& \bigg(\frac{8 \pi}{3 m^2_{\rm Pl}} \frac{\pi^2 g_{\star} T^4}{30}\bigg)^{1/2},
\label{hubble}
\end{eqnarray}
where $\rho$ is the total energy density and $m_{\rm Pl} \approx 1.22 \times 10^{19}$ GeV is the Planck mass. In the second line of this equation, we have related the energy density to the temperature of the bath, $\rho=\pi^2 g_{\star} T^4/30$, where $g_{\star}$ counts the number of effectively massless degrees-of-freedom:
\begin{equation}
g_{\star} \equiv \sum_{i={\rm bosons}} g_i \bigg(\frac{T_i}{T}\bigg)^4 + \frac{7}{8} \sum_{i={\rm fermions}} g_i \bigg(\frac{T_i}{T}\bigg)^4.
\end{equation}
Here $g_i$ is the number of internal degrees-of-freedom of state $i$. Among the particle content of the Standard Model, $g_{\star}$ varies from 106.75 at temperatures well above 100 GeV, to 10.75 at temperatures between 1 and 100 MeV, and to 3.36 at temperatures below the electron mass.

\begin{figure}[t!]
  \includegraphics[width=0.55\linewidth]{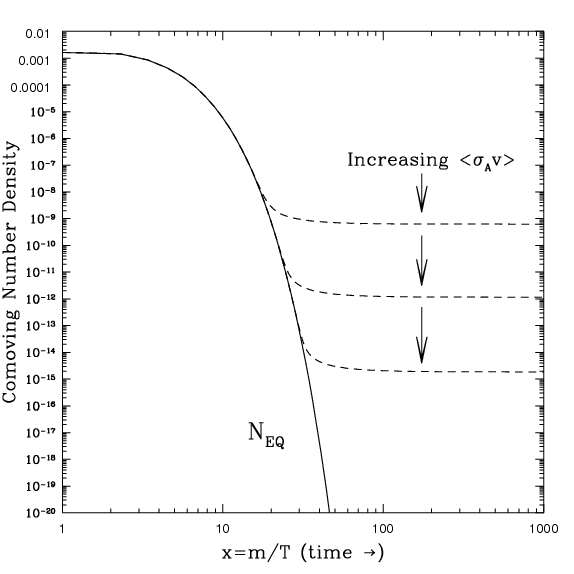}
  \caption{The freeze-out of a thermal relic. The solid line denotes the equilibrium number density, as a function of the mass of the particle divided by the temperature of the bath. The dashed lines show the number density after the relic has fallen out of equilibrium. The greater the annihilation cross section of the relic, $\langle \sigma v \rangle$, the smaller will be the relic abundance that survives the Big Bang.}
\label{fig:FreezeOut}
\end{figure}

At high-temperatures ($T \gg m_X$), the number density of a particle species $X$ that is in equilibrium is given by
\begin{eqnarray}
n^{\rm Eq}_X = \begin{cases} (\zeta(3)/\pi^2)g_XT^3 \,\,\,\,\,\,\,\,\, \,\,\,\,\,\,\,\,\,\, ({\rm Bose}) \\   (3/4) (\zeta(3)/\pi^2)g_X T^3 \, \,\,\,\, ({\rm Fermi}),  \end{cases}
\label{highT}
\end{eqnarray}
where $\zeta(3) \approx 1.20206$ and $g_X$ is the number of internal degrees-of-freedom of $X$. At low-temperatures ($T \ll m_X$), the equilibrium number density is instead given by
\begin{equation}
n^{\rm Eq}_X = g_X \bigg(\frac{m_XT}{2\pi}\bigg)^{3/2} e^{-m_X/T}.
\label{lowT}
\end{equation}
In each of these expressions for $n_X^{\rm Eq}$, we have assumed that there is no appreciable chemical potential, such as that which might arise from a primordial asymmetry between dark matter and anti-dark matter, for example. 

Unless the interactions of $X$ with the Standard Model are extremely feeble (a case I will consider later), the $X$ population will be in chemical and kinetic equilibrium with the thermal bath in the early universe. As the temperature drops below $m_X$, however, the $X$ abundance becomes exponentially suppressed (see Eq.~\ref{lowT}) until the rate of Hubble expansion exceeds that of annihilation. At that point in time, the $X$ population freezes out of equilibrium, by which I mean that its co-moving number density ($n_X a^3$, where $a$ is the scale factor) stops appreciably changing. The Hubble rate, $H$, exceeds that of the annihilation rate, $n_{X} \langle \sigma v \rangle$, when the temperature drops to $T_{\rm F}$, given as follows: 
\begin{eqnarray}
\frac{m_X}{T_{\rm F}} \approx 23 + \ln\bigg[\bigg(\frac{\sigma v}{2.2\times 10^{-26} \, {\rm cm}^3/{\rm s}}\bigg) \bigg(\frac{80}{g_{\star}}\bigg)^{1/2} \bigg(\frac{g_X}{2}\bigg)\bigg(\frac{m_X/T_{\rm F}}{23}\bigg)^{3/2} \bigg(\frac{T_{\rm F}}{10 \, {\rm GeV}}\bigg)\bigg].  
\end{eqnarray}
In other words, the $X$ population freezes out when the temperature of the universe drops to a value $\sim$\,20 times smaller than $m_X$. Although $T_{\rm F}$ is a function of the particle's annihilation cross section and number of internal degrees-of-freedom, the dependence on these quantities is only logarithmic, and $m_X/T_{\rm F} \sim 10-30$ across a wide range of values. 

After freeze-out, the total number of $X$ particles is approximately conserved, and the value of $n_X$ simply scales as $a^{-3}$ due to the expansion of the universe. The density of the $X$ population today is thus given by:
\begin{eqnarray}
\rho^{\rm today}_X &=& m_X n^{\rm today}_X \nonumber \\
&\approx& m_X n_X^{\rm Eq} (T_F) \, a^3_{\rm F},
\end{eqnarray}
where $a_{\rm F}$ is the scale factor at freeze-out. Numerically, this results in the following abundance:
\begin{eqnarray}
\Omega_X h^2 \approx 0.12 \, \bigg(\frac{2.2\times 10^{-26} \, {\rm cm}^3/{\rm s}}{\langle \sigma v\rangle}\bigg)\bigg(\frac{80}{g_{\star}}\bigg)^{1/2}\bigg(\frac{m_X/T_{\rm F}}{23}\bigg),
\end{eqnarray}
where $\Omega_X \equiv \rho_X/\rho_{\rm crit}$ is the density in terms of the critical density and $h$ is the current Hubble constant in units of 100 km/s/Mpc. For reference, cosmological measurements (including those of the cosmic microwave background) indicate that the average density of cold dark matter is near the benchmark value used in this expression, $\Omega_{\rm DM} h^2 \approx 0.11933 \pm 0.00091$~\cite{Aghanim:2018eyx}.

Note that we have assumed in this calculation that $X$ freezes out of equilibrium at a temperature well below its mass, making $X$ a cold thermal relic. If the relic is very light or feebly coupled, this may not be the case. For a particle species that freezes out when relativistic, one would repeat this calculation using Eq.~\ref{highT} to determine the abundance at freeze-out, an arriving at a very different result. Standard Model neutrinos are a well known example of a hot relic, for which this calculation yields $\Omega_{\nu+\bar{\nu}} h^2 \approx 0.0011 \, (m_{\nu}/0.1 \, {\rm eV})$. Given that the observed large scale structure of our universe rules out the possibility that any sizable fraction of the dark matter is hot, however, I will focus here on the case of dark matter in the form of a cold thermal relic.

\subsection{General Considerations Regarding the Origin of Dark Matter}

In the calculation presented above, we made a number of assumptions regarding the nature of the dark matter and its origin. In particular, we assumed that:
\begin{enumerate}
\item{$X$ is stable, or at least cosmologically long-lived.}
\item{$X$ interacts with the Standard Model strongly enough to reach equilibrium at some point in the early universe.}
\item{There are no other mechanisms that contribution to the production of $X$ particles after freeze-out.}
\item{The early universe was radiation dominated, and space expanded at the rate predicted by general relativity.} 
\end{enumerate}

Any of these conditions could plausibly be violated, of course. If the first of these conditions is not the case, however, then $X$ cannot be the dark matter, since the density of dark matter in the universe today has been measured to be similar to its abundance during the formation of the cosmic microwave background (CMB)~\cite{PalomaresRuiz:2007ry,Poulin:2016nat}. To satisfy the second condition listed above, the rate for interactions between the dark matter and the Standard Model must exceed that of Hubble expansion, $n^{\rm Eq}_X \sigma v  \gsim H$. For interactions in the form of $X$ annihilations, for example, this condition can be written as follows (for $T\gg m_X)$:
\begin{eqnarray}
n^{\rm Eq}_X \sigma v  &\gsim& H \\
\frac{a \zeta(3) g_X T^3}{\pi^2} \,  \sigma v &\gsim&  \bigg(\frac{8 \pi}{3 m^2_{\rm Pl}} \frac{\pi^2 g_{\star} T^4} {30}\bigg)^{1/2}, \nonumber
\end{eqnarray}
where $a=1$ (3/4) for the case in which $X$ is a boson (fermion). This reduces to the following condition to reach equilibrium:
\begin{equation}
\sigma v \gsim 10^{-39} \, {\rm cm}^3/{\rm s} \, \times \bigg(\frac{{\rm TeV}}{T}\bigg) \bigg(\frac{100}{g_{\star}}\bigg)^{1/2}.
\end{equation}
This is a {\it very} small cross section, many orders of magnitude smaller that that required to generate an acceptable thermal relic abundance. This ensures that any particle species with anything but the feeblest of interactions with the Standard Model will be easily maintained at equilibrium in the early universe (until freeze-out occurs). 

A stable particle species which does not interact enough to reach equilibrium with the thermal bath of Standard Model particles could be produced through a variety of mechanisms. Such possibilities include the process of thermal freeze-in (in which the $X$ particles are produced though the interactions of Standard Model particles, without the $X$ abundance ever reaching equilibrium), production through out-of-equilibrium decays~\cite{Gelmini:2006pq,Gelmini:2006pw,Merle:2015oja,Merle:2013wta,Kane:2015jia}, misalignment production (such as in the case of the QCD axion), or through the oscillations of Standard Model neutrinos into a cosmologically long-lived sterile neutrino~\cite{Dodelson:1993je,Shi:1998km,Merle:2013wta}. We also note that if the expansion history of the early universe were substantively different from that predicted in the standard radiation-dominated picture, the abundance of dark matter that emerges from the Big Bang could be altered in important ways. Examples include scenarios with an early matter dominated era~\cite{Berlin:2016gtr,Berlin:2016vnh,Gelmini:2006pq,Gelmini:2006pw} or a period of late-time inflation~\cite{Davoudiasl:2015vba}.

Taken together, these considerations force us to the conclusion that the particles that make up the dark matter must either, 1) interact at a level such that they freeze out of equilibrium to yield the measured abundance (or less, if the dark matter consists of multiple components), or 2) interact so little that they never became populated to the equilibrium abundance. Any stable particle species that interacts at a level in between these two cases will emerge from the early universe with an abundance that exceeds the measured cosmological dark matter density.  This provides us with considerable motivation to consider dark matter in the form of a particle that annihilates (or is otherwise depleted) at a rate equivalent to $\langle \sigma v \rangle \simeq 2.2 \times 10^{-26}$ cm$^3/$s at the time and temperature of thermal freeze-out. This cross section thus represents an important benchmark for indirect searches. 

Similar arguments can also allow us to place upper and lower limits on the mass of such a thermal relic. In particular, the cross section that is required to generate the measured dark matter abundance violates partial wave unitarity unless $m_X \lsim 120$ TeV~\cite{Griest:1989wd}, while the successful predictions of Big Bang Nucleosynthesis require $m_{X} \gsim (1-10)$ MeV~\cite{Boehm:2013jpa}. These two constraints provide us wth a natural range of masses for the class of dark matter candidates known as weakly interacting massive particles (WIMPs).

Although dark matter with an annihilation cross section of around $\langle \sigma v \rangle \simeq 2.2 \times 10^{-26}$ cm$^3/$s is indeed well-motivated by the above arguments, there are many viable models in which the dark matter annihilates at a higher or lower rate. In the following subsection, I will summarize some of the ways in which dark matter might be predicted to annihilate with a larger or smaller cross section in the universe today than would be expected from the simple thermal relic abundance argument described above.

\subsection{Departures From $\langle \sigma v \rangle \approx 2 \times 10^{-26}$ cm$^3/$s}

\begin{center}
{\it 1. Velocity Dependent Processes}
\end{center}

Depending on the spin of a dark matter candidate and the nature of the interactions that lead to its annihilations, the resulting cross section may or may not depend on the relative velocity between the two annihilating particles. Far from any resonances or thresholds, it is often useful to write the annihilation cross section as a Taylor series expansion in powers of $v^2$:
\begin{eqnarray}
\sigma v = a +b v^2 + c \, \mathcal{O}(v^4),
\end{eqnarray}
where $a$, $b$ and $c$ are the coefficients of this expansion. $s$-wave annihilation amplitudes contribute to all orders of this expansion, whereas $p$-wave amplitudes only contribute to the $v^2$ and higher order terms. For this reason, dark matter models which annihilate with a cross section that scales as $\sigma v \propto v^2$ are often referred to as being ``$p$-wave suppressed''. Since the velocities of dark matter particles found in halos today are generally around $v \sim 10^{-3}c$ (compared to $v \sim 0.3\, c$ at the temperature of thermal freeze-out), we expect the current annihilation rate of a $p$-wave suppressed dark matter candidate to be suppressed by a factor of roughly $\sim [10^{-3}/0.3]^2 \sim 10^{-5}$. As a result, whereas thermal relics with a velocity-independent (\ie~$s$-wave) cross section are generally excluded by current experiments and telescopes for masses up to $\mathcal{O}(10^2$ GeV), indirect detection experiments are not generally sensitive to $p$-wave suppressed dark matter candidates.

For concreteness, consider dark matter that annihilates to a pair of fermions, $f\bar{f}$, through an $s$-channel Feynman diagram. In Table~\ref{pwave}, we summarize the velocity dependance of this annihilation cross section for a variety of couplings of the mediator to the dark matter and to the final state fermions (see Ref.~\cite{Berlin:2014tja}). Of the 16 linearly independent combinations of couplings, 7 lead to a cross section that is $p$-wave suppressed ($\sigma v \propto v^2$). One should keep in mind that in many realistic dark matter models, more than one these interactions exist, leading to a combination of velocity-independent and velocity-suppressed contributions.

\begin{table*}[t]
\centering
   \begin{tabular}{| c | c | c | c | c |}
   \hline
    &\multicolumn{4}{|c|}{Fermion Bilinear} \\ \hline \hline
  Fermionic DM  & $\bar f f$ & $\bar f \gamma^5 f$ & $\bar f \gamma^\mu f$ & $\bar f \gamma^\mu\gamma^5 f$ \\ \hline  
     $\bar X  X$ & $\sigma v \sim v^2$ & $\sigma v \sim v^2$ & $-$ & $-$ \\ \hline
     $\bar X \gamma^5 X$ & $\sigma v \sim 1$&$\sigma v \sim 1$ & $-$ & $-$ \\ \hline
     $\bar X \gamma^\mu X$ & $-$ & $-$ & $\sigma v \sim 1$ & $\sigma v \sim 1$ \\ \hline
     $\bar X \gamma^\mu\gamma^5 X$ & $-$ & $-$ & $\sigma v \sim v^2$ & $\sigma v \sim 1$ \\ \hline \hline
 Scalar DM         &\multicolumn{4}{|c|}{} \\ \hline 
     $\phi^{\dagger}  \phi$ & $\sigma v \sim 1$ & $\sigma v \sim 1$ & $-$ & $-$  \\ \hline
     $\phi^{\dagger}  \overset{\leftrightarrow}{\partial_{\mu}}  \phi$ & $-$ & $-$ & $\sigma v \sim v^2$ & $\sigma v \sim v^2$ \\ \hline \hline
  %
  Vector DM         &\multicolumn{4}{|c|}{} \\ \hline 
     $X^\mu X_\mu^{\dagger}$ & $\sigma v \sim 1$ & $\sigma v \sim 1$ & $-$ & $-$ \\ \hline
 $X^\nu \partial_\nu X_\mu^{\dagger}$ & $-$ & $-$ & $\sigma v \sim v^2$ &  $\sigma v \sim v^2$ \\  \hline     
     \end{tabular}
\caption{A summary of the velocity dependance of the dark matter annihilation cross section for $s$-channel diagrams to fermion-antifermion final states. Of the 16 linearly independent combinations of couplings shown here, 7 lead to a cross section that is $p$-wave suppressed ($\sigma v \propto v^2$).}     
\label{pwave}
\end{table*}

\begin{center}
{\it 2. Resonant Annihilations}
\end{center}

If the dark matter annihilates through or near a resonance, its cross section could be much higher or lower during freeze-out than at the very low velocities found in halos today~\cite{Griest:1990kh,Hooper:2013qjx}. Consider, for example, an annihilation cross section of the following form:
\begin{equation}
\sigma v = \frac{\alpha^2 s}{(M^2_{\rm med}-s)^2+M^2_{\rm med}\Gamma^2_{\rm med}},
\end{equation}
where $M_{\rm med}$ and $\Gamma_{\rm med}$ are the mass and the width of the particle mediating the annihilation process and $\alpha^2$ normalizes the cross section. The Mandelstam variable, $s=4m^2_{X}/(1-v^2)$, is equal to $s_{v \rightarrow 0} = 4 m^2_X$ in the low-velocity limit, and to a value roughly 10\% larger at the temperature of thermal freeze-out, $s_{\rm FO} \simeq 4m^2_X/(1-0.1) \simeq 1.1 (4 m^2_X)$. As a first case, consider a scenario in which $M_{\rm Med} \simeq 2 m_X$, enabling the dark matter to annihilate resonantly at low velocities. In the narrow width approximation ($\Gamma_{\rm med} \ll M_{\rm med}$), this leads to an {\it enhancement} of the low-velocity cross section by a factor of $\sim 8 \times [(M_{\rm med}/\Gamma_{\rm med})/30]^2$ relative to the velocity-independent case. Alternatively, we could instead consider a case in which $M_{\rm Med} \simeq 2.1 m_X$, for which the dark matter annihilates on resonance during freeze-out. In this case, the low-velocity cross section is {\it suppressed} by a similar factor.

\begin{center}
{\it 3. Coannihilations}
\end{center}

Instead of being depleted through self-annihilations, the dark matter abundance could instead be established through coannihilations with another particle species, $X'$~\cite{Griest:1990kh,Edsjo:1997bg,Ellis:1998kh}. The relative abundance of such a state at the temperature of freeze-out can roughly be estimated as follows:
\begin{eqnarray}
\frac{n_{X'}}{n_X} &\sim& \frac{e^{-m_{X'}/T_{\rm F}}}{e^{-m_{X}/T_{\rm F}}} \\
&=& e^{-\Delta m_X/T_{\rm F}} \nonumber \\
&\sim& e^{-20 \Delta},\nonumber
\end{eqnarray}
where $\Delta \equiv (m_{X'}-m_X)/m_X$ is the fractional mass splitting between the two states. For large mass splittings ($\Delta \gg 0.1$), $n_{X'} \ll n_{X}$, and the $X'$ population will play little role in the process of thermal freeze-out, or in determining the final $X$ abundance. For smaller splittings ($\Delta \lsim 0.1$), however, a significant number of $X'$ particles will be present during freeze-out, potentially assisting in the depletion of the $X$ population.

To calculate the impact of coannihilations on the thermal relic abundance, we introduce the following effective cross section:
\begin{equation}
\sigma_{\rm eff}(T) \equiv \sum_{i,j} \sigma_{i,j} \frac{g_i g_j}{g^2_{\rm eff} (T)} (1+\Delta_i)^{3/2} (1+\Delta_j)^{3/2} e^{-m_X(\Delta_i+\Delta_j)/T},
\end{equation}
where $T$ is the temperature and $g_{i,j}$ and $\Delta_{i,j}$ are the number of internal degrees-of-freedom and the fractional mass splittings (relative to that of $X$) of state $i$, and 
\begin{equation}
g_{\rm eff}(T) \equiv \sum_i g_i (1+\Delta_i)^{3/2} e^{-m_X\Delta_i/T}.
\end{equation}
As an example, consider two states ($X$ and $X'$) that are nearly degenerate ($\Delta_{X'} \ll 1$) and that have an equal number of internal degrees-of-freedom ($g_X=g_{X'}$).  In this case, the effective annihilation cross section reduces to $\sigma_{\rm eff} \simeq 0.5 \sigma_{XX} + 0.5 \sigma_{X'X'} +\sigma_{XX'}$. If $\sigma_{XX'} \gsim \sigma_{XX}$, coannihilations will play a major role in the depletion of the $X$ abundance. In the opposite case ($\sigma_{XX'} \ll \sigma_{XX}, \sigma_{X'X'}$), the $X$ and $X'$ populations each freeze out and contribute to the final dark matter abundance independently. 

\begin{center}
{\it 4. Asymmetric Dark Matter}
\end{center}

If you were to carry out the calculation of the thermal relic abundance as described above for the case of protons and electrons, you would find that almost no such particles should survive the conditions of the early universe. The baryon-antibaryon annihilation cross section is much larger than that needed to yield a cosmologically interesting abundance. The abundance of baryons found in our universe is instead determined by the presence of a primordial matter-antimatter asymmetry. Namely, for reasons that are not yet understood, the early universe contained slightly more baryons than antibaryons (and more quarks than antiquarks prior to the QCD phase transition). These particles stopped annihilating not when the expansion rate caused their abundance to freeze out, but instead when annihilations had destroyed all of antibaryons that had once been present in the universe.
 
It is possible that there could have also been a primordial asymmetry between the number of dark matter particles, $X$, and antiparticles, $\bar{X}$, in the early universe~\cite{Zurek:2013wia,Graesser:2011wi,Lin:2011gj,Iminniyaz:2011yp}. If this is the case, then the $X\bar{X}$ annihilation cross section could in principle be much larger than our benchmark value of $2\times 10^{-26}$ cm$^3/$s. But despite this large cross section, the annihilation rate could still be very low in the universe today, as a result of the absence of $\bar{X}$ particles. In such a scenario, the prospects for indirect detection could be highly suppressed. Alternatively, $X-\bar{X}$ oscillations could repopulate this population, potentially leading to very high annihilation rates in the current epoch~\cite{Cirelli:2011ac,Buckley:2011ye}.

\begin{center}
{\it 5. Sommerfeld Enhancements}
\end{center}

In some dark matter models, long-range interactions can enhance the annihilation cross section at low velocities~\cite{Hisano:2004ds,ArkaniHamed:2008qn}. This effect, known as the ``Sommerfeld enhancement'', is most pronounced in cases in which the mediator is much lighter than the dark matter itself, $M_{\rm med} \lsim m_X v$. A well studied example is that of dark matter in the form of a TeV-scale, wino-like neutralino. In this case, the low-velocity annihilation cross section can exceed the thermal relic benchmark value by up to 1 to 2 orders of magnitude.

\begin{center}
{\it 6. Out of Equilibrium Decays and Other Non-Thermal Production Mechanisms}
\end{center}

In addition to any thermal abundance of a particle species that might arise, an additional non-thermal population could be generated through, for example, the decays of another species that is not in equilibrium with the thermal bath~\cite{Gelmini:2006pq,Gelmini:2006pw,Merle:2015oja,Merle:2013wta,Kane:2015jia}. Moduli are an example of a theoretically well-motivated state that is predicted to fall out of equilibrium before it decays, potentially leading to the production of a non-thermal dark matter population. In such scenarios, it is possible for the dark matter annihilation cross section to be considerably higher than generally predicted for a thermal relic.

\begin{center}
{\it 7. Non-Standard Cosmological Histories}
\end{center}

Unless altered by new physics, the energy density of our universe was dominated by radiation (\ie~relativistic particles) during the first $\sim$\,$10^5$ years of its (post-inflationary) history. If there exists a long-lived particle species that becomes non-relativistic in the early universe, the energy density of its population will evolve like $\rho \propto a^{-3}$, whereas radiation dilutes as $\rho \propto a^{-4}$. As a result, the non-relativistic species will increasingly come to dominate the energy density of the early universe, potentially leading to an era of matter domination~\cite{Fornengo:2002db,Gelmini:2006pq,Kane:2015jia,Berlin:2016gtr,Berlin:2016vnh,Gelmini:2006pw}. This could impact the abundance of dark matter in at least two different ways. First of all, when the long-lived particles ultimately decay, they could produce dark matter particles, as described in the paragraph directly above this one. Furthermore, such decays could dilute the dark matter's thermal abundance, lowering the annihilation cross section that is required to generate the measured cosmological density.  Alternatively, the expansion history of the early universe could be altered by the presence of an era of rapid expansion, known as late-time inflation~\cite{Lyth:1995ka,Cohen:2008nb,Boeckel:2011yj,Boeckel:2009ej,Davoudiasl:2015vba}. If this occurs after freeze-out, such an event would dilute the abundance of dark matter and lower the expected annihilation cross section.

\section{Gamma-Ray Searches for Dark Matter Annihilation Products}
\label{gammasec}

If the dark matter annihilates with a cross section near the thermal relic benchmark value, $\langle \sigma v \rangle \simeq 2.2\times 10^{-26}$ cm$^3/$s, this could potentially lead to an observable flux of energetic particles, including gamma rays and cosmic rays. Searches for dark matter using gamma-ray telescopes benefit from the fact that these particles are not deflected by magnetic fields and are negligibly attenuated over Galactic distance scales, making it possible to acquire both spectral and spatial information, unmolested by astrophysical effects.

The possibility that gamma-ray telescopes could be used to detect the annihilation products of dark matter particles was first suggested in a pair of papers published in 1978 by Jim Gunn, Ben Lee, Ian Lerche, David Schramm and Gary Steigman~\cite{1978ApJ...223.1015G}, and by Floyd Stecker~\cite{1978ApJ...223.1032S}. Today, four decades later, gamma-ray searches for dark matter provide us with some of the most stringent and robust constraints on the dark matter's annihilation cross section.

The dark matter annihilation rate per volume is given by $\langle \sigma v \rangle \, \rho_X^2/2m^2_X$, where $\rho_X$ is the dark matter density and the factor of 1/2 is included to avoid double counting the annihilations of particle A with particle B, and particle B with particle A. Here we are assuming that the annihilating particles are their own antiparticle ($XX$). If  we were instead to consider annihilations between dark matter and anti-dark matter ($X\bar{X}$), the annihilation rate would be half as large for a given value of the cross section. However, the annihilation cross section must also be twice as large in this case in order to obtain the desired relic abundance, and thus the overall annihilation rate of a thermal relic today remains the same, regardless of whether the dark matter candidate is or is not its own antiparticle.

To calculate the spectrum and angular distribution of gamma rays from dark matter annihilations per unit time from within a solid angle, $\Delta \Omega$, we integrate the annihilation rate over the solid angle observed, and over the line-of-sight:
\begin{eqnarray}
\Phi_{\gamma} (E_\gamma, \Delta \Omega) = \frac{1}{2}  \frac{dN_{\gamma}}{dE_{\gamma}} \frac{ \langle \sigma v \rangle}{4\pi m^2_{X}} \int_{\Delta \Omega} \int_{los} \rho_X^2(l,\Omega) dl d\Omega,
\label{gamma}
\end{eqnarray}
where $dN_{\gamma}/dE_{\gamma}$ is the spectrum of gamma rays produced per annihilation, which depends on the mass of the dark matter particle and on the types of particles that are produced in this process. In practice, such spectra are often calculated using software such as PYTHIA~\cite{Sjostrand:2006za}. In addition to prompt gamma rays, dark matter annihilations can produce electrons and positrons which generate gamma rays through inverse Compton and bremsstrahlung processes~\cite{Belikov:2009cx,Profumo:2009uf,Cirelli:2013mqa}. 

The basic characteristics of $dN_{\gamma}/dE_{\gamma}$ depend primarily on the dominant annihilation channels of the dark matter particle. For annihilations to quark-antiquark pairs, the resulting jets produce photons through the decays of neutral pions, resulting in a spectrum that typically peaks at an energy around $\sim$\,$m_X/20$ (in $E_{\gamma}^2 dN_{\gamma}/dE_{\gamma}$ units). For dark matter that is heavy enough to produce $W$ or $Z$ pairs in their annihilations, the resulting gamma-ray spectrum is similar. In contrast, if the dark matter annihilates to charged lepton pairs, the resulting spectrum is predicted to be quite different. Annihilations to $\tau^+ \tau^-$ produce a gamma-ray spectrum that is fairly sharply peaked around $\sim$\,$m_X/3$ (due to the harder spectrum of neutral pions). In the case of annihilations to $e^+ e^-$ or $\mu^+ \mu^-$ the gamma-ray spectrum is dominated by final state radiation (rather than pion decay) and inverse Compton scattering, generally resulting in a smaller flux of higher-energy photons.

The quantity described by the integrals in Eq.~\ref{gamma} is often referred to as the $J$-factor, which encodes all of the relevant astrophysical information. To build some intuition for the annihilation $J$-factor, consider the simple example of dark matter particles annihilating in a spherical dwarf galaxy of radius $r$, uniform density $\rho$, and located at a distance $d$. For $d \gg r$, this $J$-factor is given by:
\begin{eqnarray}
J \equiv \int_{\Delta \Omega} \int_{los} \rho_X^2(l,\Omega) dl d\Omega \simeq \frac{4 \pi r^3 \rho^2_X}{3d^2}.
\end{eqnarray}
From this simple example, we can see that the most promising targets of gamma-ray searches for dark matter are those that:
\begin{enumerate}
\item{Have a high density of dark matter ($J \propto \rho_X^2$)}
\item{Are nearby ($J \propto d^{-2}$)}
\item{Are extended across a large volume ($J \propto V$)}
\item{Are accompanied by low and/or well-understood astrophysical backgrounds.}
\end{enumerate}

The first three of these conditions are best satisfied by the inner volume of the Milky Way, sometimes referred to as the Galactic Center. The Galactic Center is almost certain to be the brightest single source of dark matter annihilation products on the sky. This direction, however, is also plagued by large and imperfectly understood astrophysical backgrounds. At the other extreme are the Milky Way's dwarf galaxies, which have much smaller $J$-factors than the Galactic Center, but are accompanied by much smaller gamma-ray backgrounds. Intermediate strategies include observations of other promising targets, including galaxy clusters~\cite{Lisanti:2017qlb}, the halo of the Milky Way, and the isotropic gamma-ray background~\cite{Ando:2015qda,Cholis:2013ena,Ackermann:2015tah,DiMauro:2015tfa,Ajello:2015mfa}.

Modern gamma-ray astronomy is conducted using a combination of space-based and ground-based telescopes, each of which offer various advantages and disadvantages. At energies between 0.1 and 100 GeV, this field is dominated by the Fermi Gamma-Ray Space Telescope, which has been in orbit around Earth since 2008. Fermi observes the entire sky with an angular resolution on order of a degree and an energy resolution of around 10\%. Among other science goals, Fermi was designed to offer unprecedented sensitivity to dark matter annihilation products~\cite{Gehrels:1999ri}, in particular from the direction of the center of the Milky Way~\cite{Bergstrom:1997fj,Berezinsky:1994wva,Gondolo:1999ef,Ullio:2001fb,Cesarini:2003nr,Peirani:2004wy,Dodelson:2007gd}. At higher energies, ground-based air Cherenkov telescopes offer the greatest sensitivity, including as HESS~\cite{Abdallah:2016ygi,Abdalla:2018mve,Abdallah:2018qtu}, VERITAS~\cite{Archambault:2017wyh,Zitzer:2015eqa} and MAGIC \cite{Ahnen:2017pqx,Ahnen:2016qkx} (and in the future, CTA~\cite{Consortium:2010bc}). While these instruments have far greater angular resolution than Fermi, they must be pointed at specific targets and are only sensitive to gamma rays with energies above around $\sim$\,$10^2$ GeV. The HAWC telescope is also sensitive in the case of very heavy dark matter particles~\cite{Abeysekara:2017jxs,Blanco:2017sbc}.

\subsection{Dwarf Galaxies}

The Milky Way's dark matter halo contains a large number of smaller subhalos, the largest of which contain stars and constitute satellites of our galaxy. The satellite population of the Milky Way includes the classical dwarfs (Draco, Ursa Minor, Sculptur, Fornax, etc.), as well as several dozen ultra-faint galaxies that were discovered using data from modern surveys, including the Sloan Digital Sky Survey (SDSS) and the Dark Energy Survey (DES)~\cite{Bechtol:2015cbp,Koposov:2015cua}.

Although dwarf galaxies are typically discovered using photometric data, spectroscopic follow up observations can measure the line-of-sight velocities of the brightest stars in these systems. This information can then be used to infer information about the underlying dark matter distribution, and to estimate the annihilation $J$-factor of each dwarf.

In making such estimates, most groups assume that a given dwarf galaxy is, 1) in steady state, 2) spherically symmetric, and 3) negligibly rotationally supported. Under such assumptions, one can derive the 2nd order Jeans equation, which can be solved and projected along the line-of-sight to produce the predicted velocity dispersion as a function of angular radius. This is then compared to the measured distribution of velocities to generate constraints on the $J$-factor of the dwarf~\cite{Martinez:2013els,Bonnivard:2014kza,Bonnivard:2015vua}.

There are a number of challenges involved in deriving constraints on dwarf galaxy $J$-factors. First of all, it is not obvious that the three assumptions mentioned in the previous paragraph are valid. In particular, dwarf galaxies are not expected to be perfect spherical, a factor which can non-negligibly skew the value of the inferred $J$-factor.  Furthermore, for many ultra-faint dwarfs, spectroscopic measurements exist for only a small number of stars. Making this more perilous is the fact that it is not always clear which stars are in fact gravitationally bound to a given dwarf. In the case of Segue 1, for example, quite different $J$-factor determinations can result depending on how the question of stellar membership is precisely treated~\cite{Bonnivard:2015vua}.

\begin{figure*}
\centering
\includegraphics[width=0.95\textwidth,clip=true]{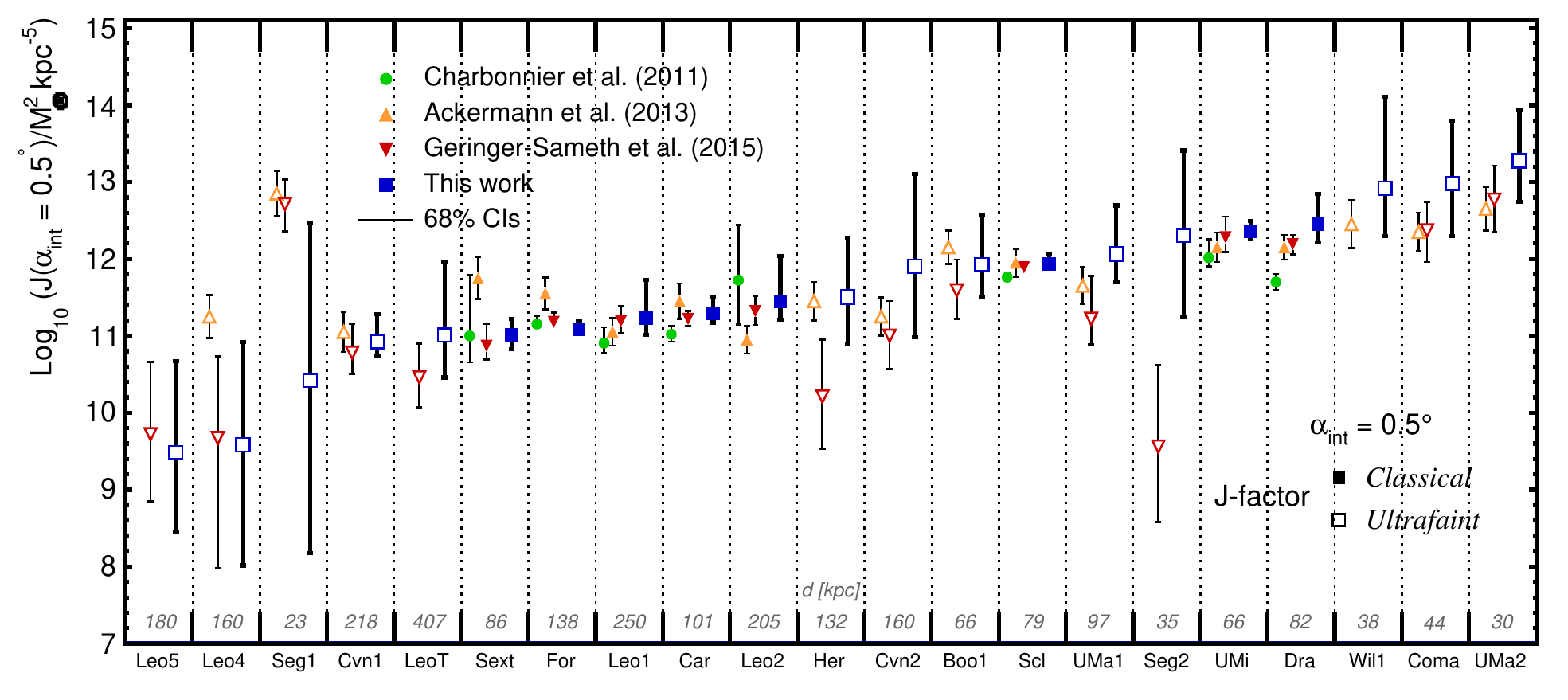} 
\caption{The $J$-factors inferred for a collection of Milky Way dwarf galaxies, averaged over a $0.5^{\circ}$ radius, as presented in Ref.~\cite{Bonnivard:2015xpq}, and compared to the results of three other groups~\cite{Charbonnier:2011ft,Geringer-Sameth:2014yza,Ackermann:2013yva}. Notice that the distance to the dwarf galaxy (shown across the bottom of the panel) is a fairly strong predictor of its $J$-factor. }
\label{Jfactors}
\end{figure*}

In Fig.~\ref{Jfactors} (from Ref.~\cite{Bonnivard:2015xpq}), we show the $J$-factor determinations for 21 Milky Way dwarf galaxies, as presented by several groups~\cite{Bonnivard:2015xpq,Charbonnier:2011ft,Geringer-Sameth:2014yza,Ackermann:2013yva}. Note that the largest $J$-factors are generally found for those dwarfs that are most nearby, and that the error bars associated with ultra-faint dwarfs are typically much larger than those of the classical dwarfs.

As an example, consider the classical dwarf galaxy Draco, which has a measured $J$-factor of $J \simeq 10^{18.8}$ GeV$^2/$cm$^5$. Combining this number with Eq.~\ref{gamma} leads to the following estimate for the gamma-ray flux from this satellite:
\begin{eqnarray}
\label{draco}
\Phi_{\gamma} &\simeq& \frac{1}{2}   \frac{ \langle \sigma v \rangle}{4\pi m^2_{X}} \, J \, \int \frac{dN_{\gamma}}{dE_{\gamma}} dE_{\gamma} \\
&\approx& 5 \times 10^{-12} \, {\rm cm}^{-2} {\rm s}^{-1} \bigg(\frac{\langle \sigma v \rangle}{2\times 10^{-26} \, {\rm cm}^3/{\rm s}}\bigg) \bigg(\frac{\int \frac{dN_{\gamma}}{dE_{\gamma}} dE_{\gamma}}{10}\bigg)\bigg(\frac{100 \, {\rm GeV}}{m_X}\bigg)^2 \bigg(\frac{J}{10^{18.8}\, {\rm GeV}^2/{\rm cm}^5}\bigg). \nonumber
\end{eqnarray}
Multiplying the above flux by Fermi's effective area of $\simeq$\,8500 cm$^2$, and by the fact that this telescope observes a given portion of the sky $\sim$\,20\% sky of the time, we arrive at an estimate that this instrument would detect approximately 0.3 photons per year from dark matter annihilations in Draco (for the parameters shown in the brackets). Given that this is much smaller than the flux associated with the extragalactic gamma-ray background (in addition to the contribution from diffuse emission mechanisms in the Milky Way), we conclude that Fermi is not sensitive to dark matter annihilation in Draco, at least for this choice of parameters. More optimistically, we could instead consider dark matter in the form of a thermal relic with a mass of 30 GeV, increasing the predicted gamma-ray flux by more than an order of magnitude. Over ten years of observation, one would expect such a scenario to lead to a few dozen signal events, which could constitute a modest excess ($\sim$\,1-2$\sigma$) over known backgrounds.

\begin{figure*}[t]
  \includegraphics[width=0.47\linewidth]{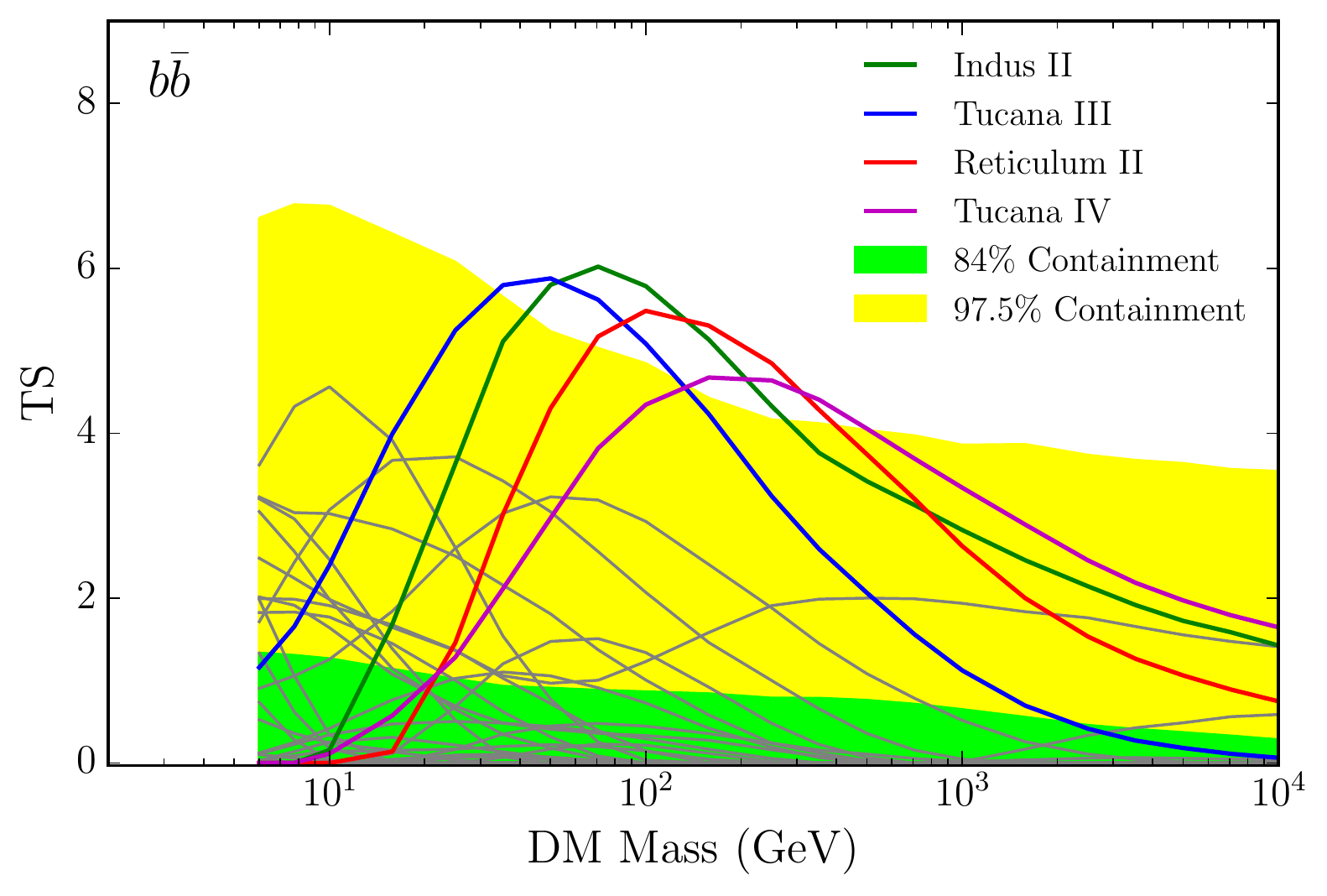}
  \includegraphics[width=0.47\linewidth]{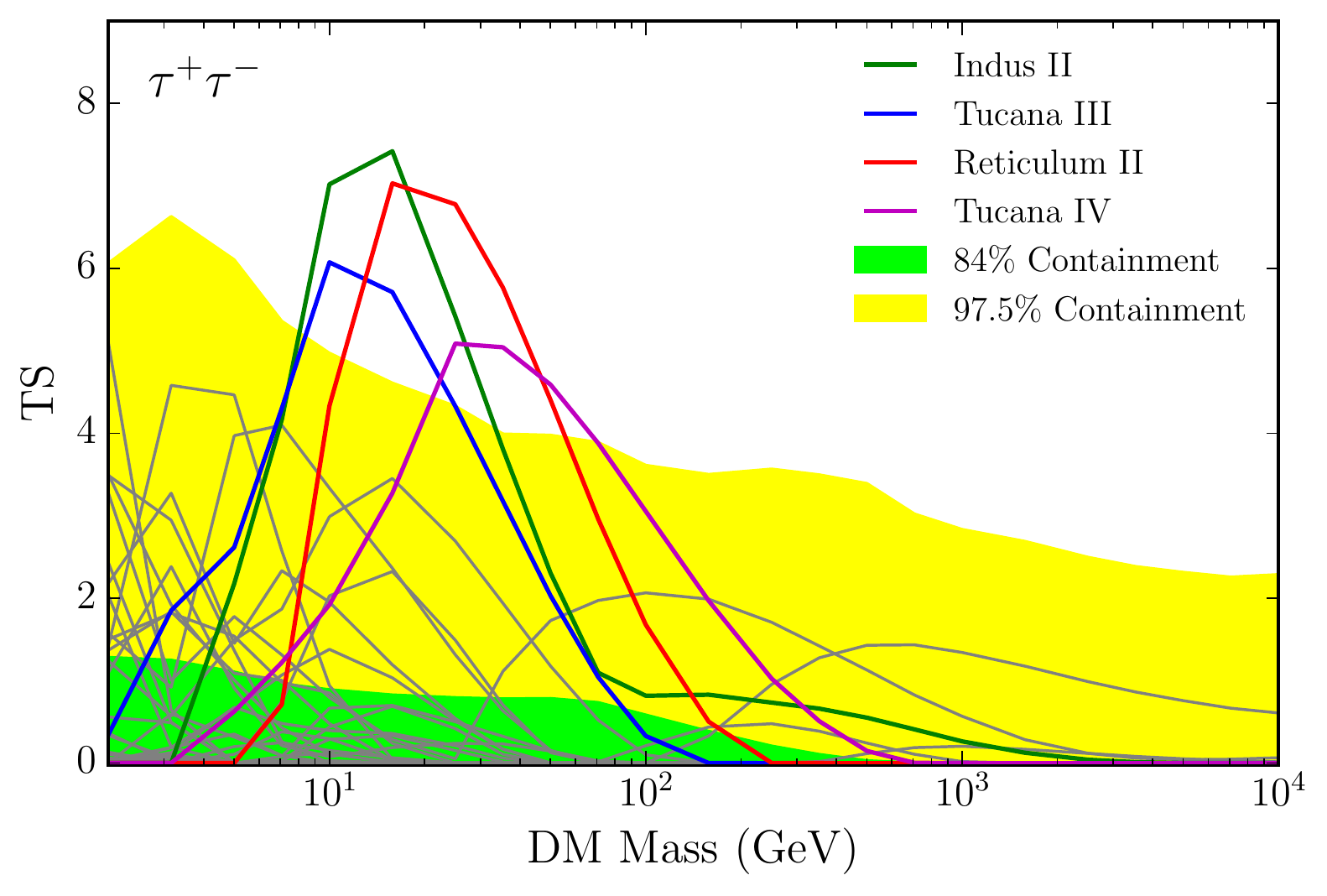}
    \includegraphics[width=0.49\linewidth]{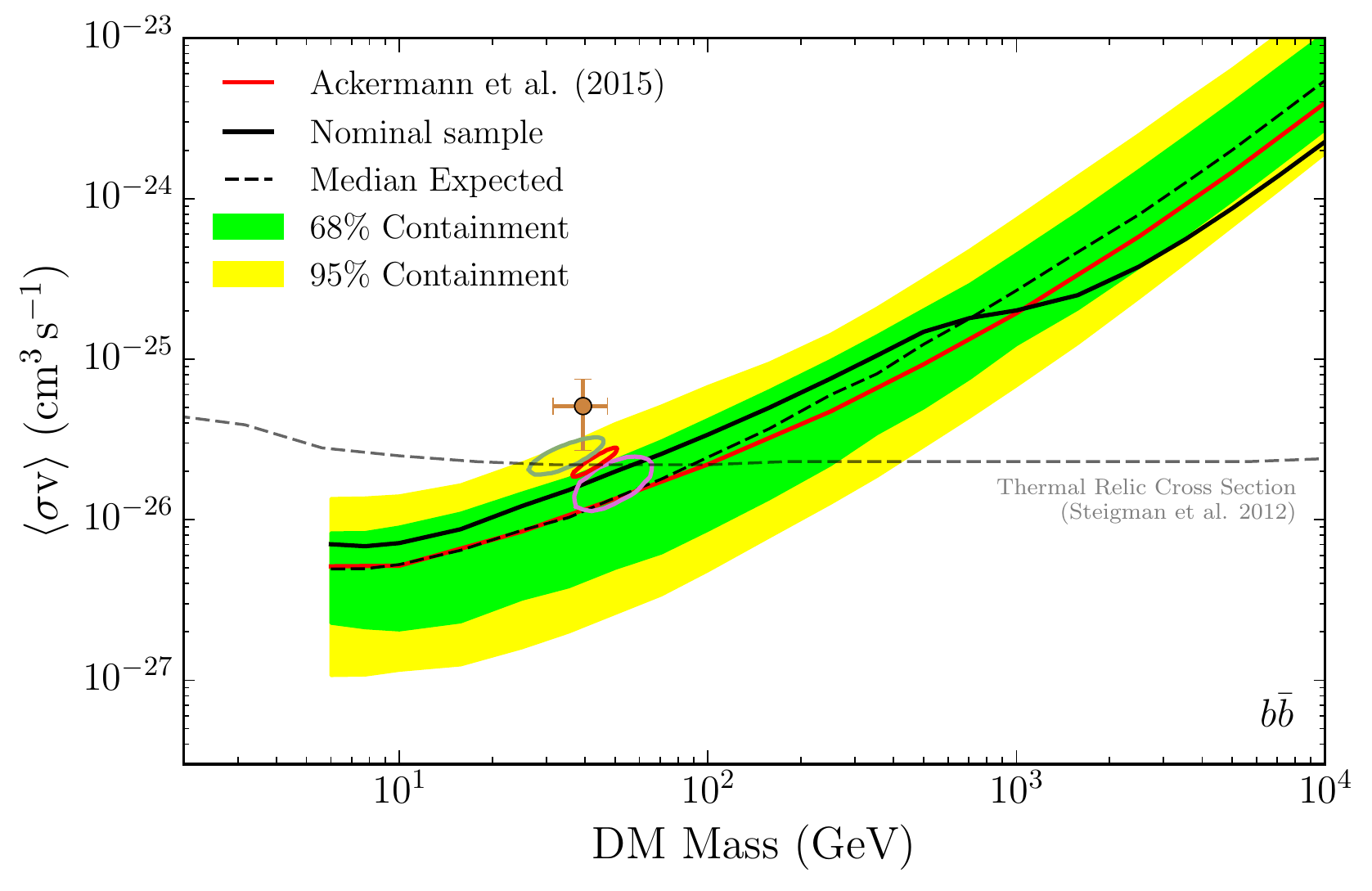}
  \includegraphics[width=0.49\linewidth]{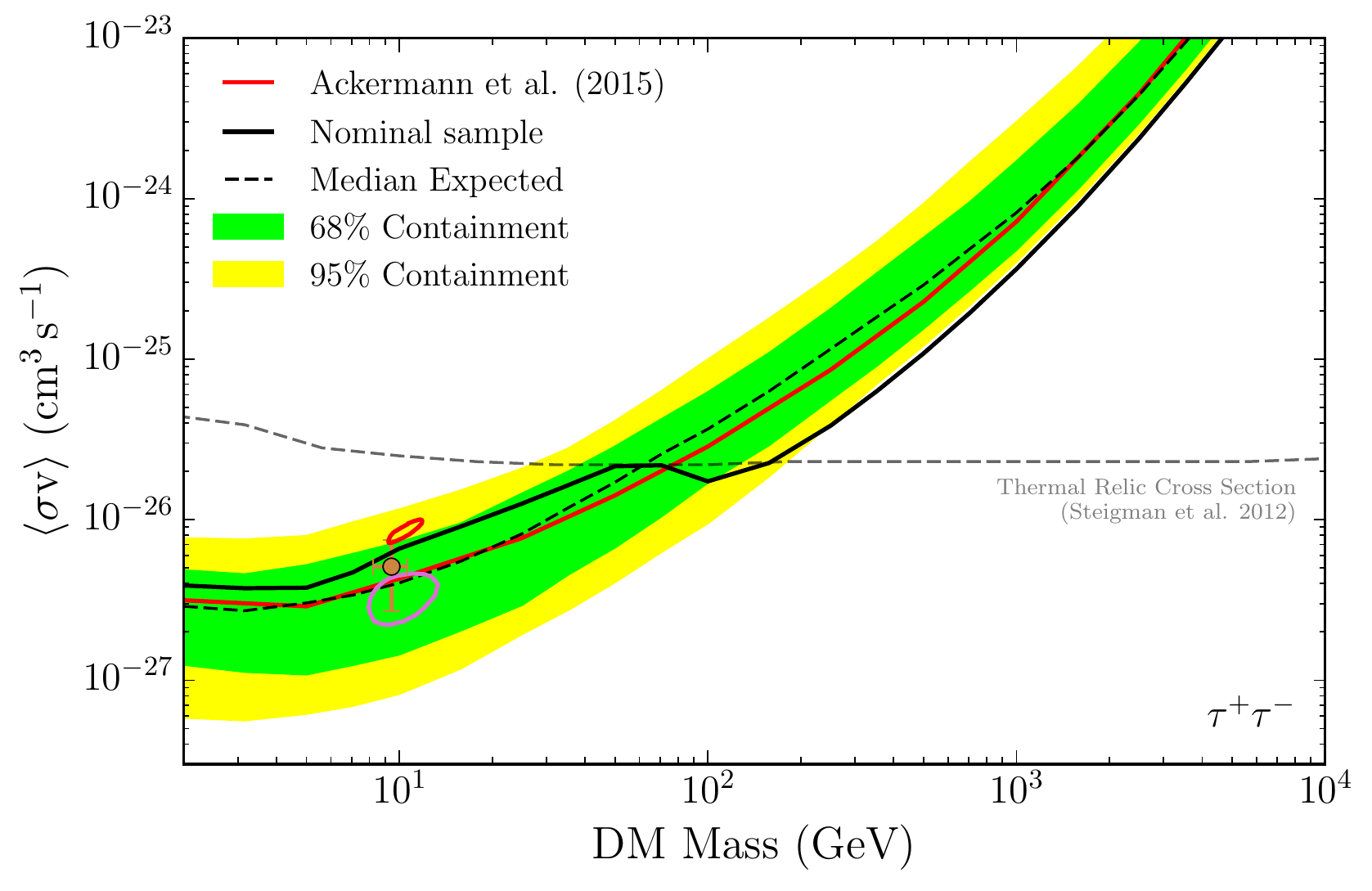}
  \caption{\label{fig:TSvsMass} Upper Frames: The local detection significance, expressed as a log-likelihood test statistic (TS), from the Fermi Collaboration's analysis of a collection of dwarf galaxies, for the cases of dark matter annihilating to $b\bar{b}$ (left) or $\tau^+ \tau^-$ (right). Lower Frames: Upper limits (95\% confidence level) on the dark matter annihilation cross section, compared to the expected sensitivity (colored bands). This analyses rules out the simple relic benchmark annihilation cross section ($\langle \sigma v \rangle \sim 2\times 10^{-26}$ cm$^3/$s) for masses up to $\sim$\,60 GeV (for the case of annihilations to $b\bar{b}$). From Ref.~\cite{Fermi-LAT:2016uux}.}
\end{figure*}

In practice, constraints are placed by stacking many dwarf galaxies as a part of a combined analysis. Some of the main results from the Fermi Collaboration's most recent dwarf galaxy analysis are shown in Fig.~\ref{fig:TSvsMass} (see also Refs.~\cite{Geringer-Sameth:2014qqa,Ackermann:2015zua,Abramowski:2014tra,Ahnen:2017pqx,Ahnen:2016qkx,Archambault:2017wyh}). This analysis is based on a stack of 15 dwarfs, and it excludes dark matter candidates with $\langle \sigma v \rangle = 2\times 10^{-26}$ cm$^3/$s up to masses of $\sim$\,60 GeV for the case of annihilations to $b\bar{b}$. It is also interesting to note that statistically modest gamma-ray excesses have been detected from the directions of a few dwarf galaxies, including Recticulum II and Tucana III~\cite{Geringer-Sameth:2015lua,Drlica-Wagner:2015xua,Hooper:2015ula}. If these are authentic signals of dark matter, it would suggest a mass in the range of $\sim 50-100$ GeV (for annihilations to $b\bar{b}$) and a cross section near $\sim$\,$10^{-26}$ cm$^3/$s.

Looking forward, the constraints on annihilating dark matter based on gamma-ray observations of dwarf galaxies are expected to improve due to, 1) the growing data set from Fermi (and future gamma-ray telescopes, such as AMIGO or e-ASTROGAM), and 2) the discovery of new ultra-faint dwarf galaxies that are expected from LSST and other surveys. It is anticipated that Fermi's sensitivity to dark matter annihilation in dwarf galaxies will improve substantially in the LSST era.

\subsection{The Galactic Center}

The Galactic Center is expected to be the single brightest source of dark matter annihilation products on the sky, but is also plagued by bright and imperfectly understood astrophysical backgrounds. Furthermore, the prospects for detecting dark matter from this region depend critically on the distribution of dark matter in the central volume of the Milky Way. In fact, the flux of dark matter annihilation products that is predicted from the innermost degree or so around the Galactic Center (corresponding to approximately the angular resolution of Fermi's Large Area Telescope) can vary by orders of magnitude, depending on the halo profile that is adopted~\cite{Hooper:2012sr,Gomez-Vargas:2013bea}. The sensitivity of ground-based gamma-ray telescopes, with much greater angular resolution, can depend even more strongly on the halo profile's inner slope~\cite{Abramowski:2011hc,Aharonian:2006wh,Silverwood:2014yza,Pierre:2014tra}.

Numerical simulations of cold, collisionless dark matter particles yield profiles with high central densities~\cite{Navarro:2008kc,Diemand:2008in}. A common parameterization for this distribution is the generalized Navarro-Frenk-White (NFW) halo profile~\cite{Navarro:1995iw,Navarro:1996gj}:
\begin{equation}
\rho( r)\propto \frac{(r/R_s)^{-\gamma}}{(1 + r/R_s)^{3-\gamma}}, 
\label{gennfw}
\end{equation}  
where $R_s \sim 20$ kpc is the scale radius of the Milky Way. While the canonical NFW profile is defined such that $\gamma=1$, other values for the inner slope are also commonly adopted (as well as other parameterizations, such as the Einasto profile~\cite{Springel:2008cc}). In particular, modern simulations which include the effects of baryonic processes have been found to yield a wide range of inner profiles, $\gamma \sim 0.5-1.4$~\cite{Gnedin:2011uj,Gnedin:2004cx,Governato:2012fa,Kuhlen:2012qw,Weinberg:2001gm,Weinberg:2006ps,Sellwood:2002vb,Valenzuela:2002np,Colin:2005rr,Scannapieco:2011yd,Calore:2015oya,Schaller:2014uwa,DiCintio:2014xia,DiCintio:2013qxa,Schaller:2015mua,Bernal:2016guq}. Empirically speaking, we have only a modest degree of information about the shape of the Milky Way's dark matter halo profile. More specifically, although many groups have presented dynamical evidence in support of dark matter's presence in the Milky Way~\cite{Weber:2009pt,Catena:2009mf,Iocco:2011jz,Bovy:2012tw,Garbari:2012ff,Bovy:2013raa,Read:2014qva,Iocco:2015xga,Pato:2015dua}, these measurements provide relatively little information about dark matter in the innermost kiloparsecs of the Galaxy.  We also note that although dark matter halos are expected to exhibit some degree of triaxiality (see, for example, Ref.~\cite{Kuhlen:2007ku}), the Milky Way's dark matter halo is generally predicted to produce an annihilation signal that is approximately radially symmetric with respect to the Galactic Center~\cite{Bernal:2016guq}.

As a simple example, consider dark matter that is distributed according to a standard NFW profile with $R_s=20$ kpc and a local density of 0.4 GeV/cm$^3$. Using Eq.~\ref{gamma}, this yields the following flux of gamma-ray annihilation products originating from the innermost 2 kpc around the Galactic Center:
\begin{eqnarray}
\label{draco}
\Phi_{\gamma} \sim 10^{-8} \, {\rm cm}^{-2} {\rm s}^{-1} \bigg(\frac{\langle \sigma v \rangle}{2\times 10^{-26} \, {\rm cm}^3/{\rm s}}\bigg) \bigg(\frac{\int \frac{dN_{\gamma}}{dE_{\gamma}} dE_{\gamma}}{10}\bigg)\bigg(\frac{100 \, {\rm GeV}}{m_X}\bigg)^2. 
\end{eqnarray}

The first thing to notice about this flux is that it is more than three orders of magnitude larger than that predicted from the brightest dwarf galaxies. The problem, of course, is that of astrophysical backgrounds. The dominant gamma-ray backgrounds from this region of the sky consist of diffuse emission resulting from, 1) pion production via cosmic-ray proton scattering with gas, 2) cosmic-ray electron scattering with radiation via inverse Compton scattering, and 3) cosmic-ray electron scattering with gas via Bremsstrahlung. Models for these backgrounds are built using inputs such as gas maps, and models of cosmic-ray transport. And while such models are often capable of describing the broad features of the observed Galactic diffuse emission, they cannot (and should not be expected to) account for the detailed spectral or morphological characteristics of this background. In addition, significant backgrounds also arise from gamma-ray point sources, such as supernova remnants, pulsars, blazars and the Milky Way's central supermassive black hole (Sgr A$^*$).

\begin{figure*}
\centering
\includegraphics[width=0.49\textwidth,clip=true]{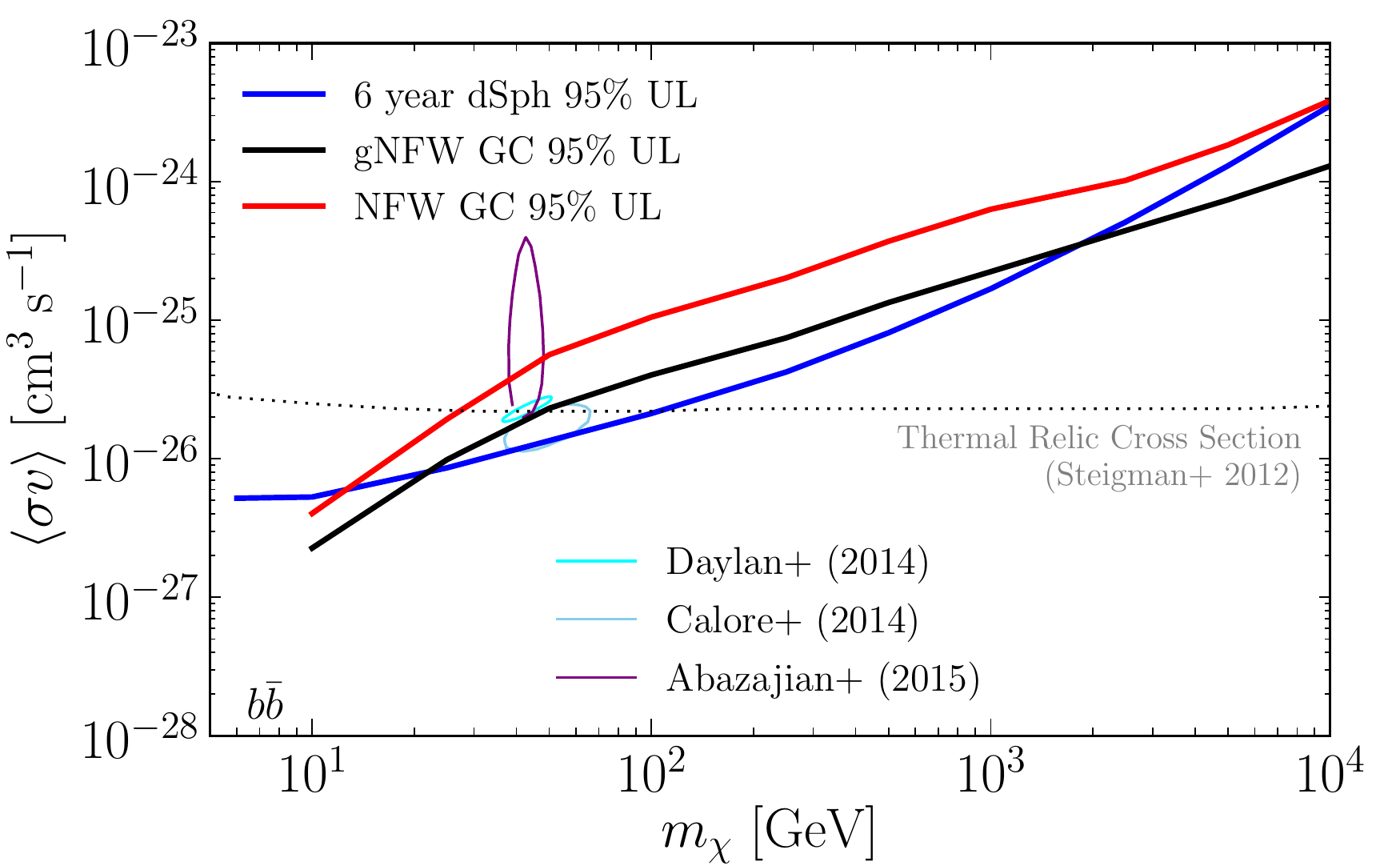} 
\includegraphics[width=0.49\textwidth,clip=true]{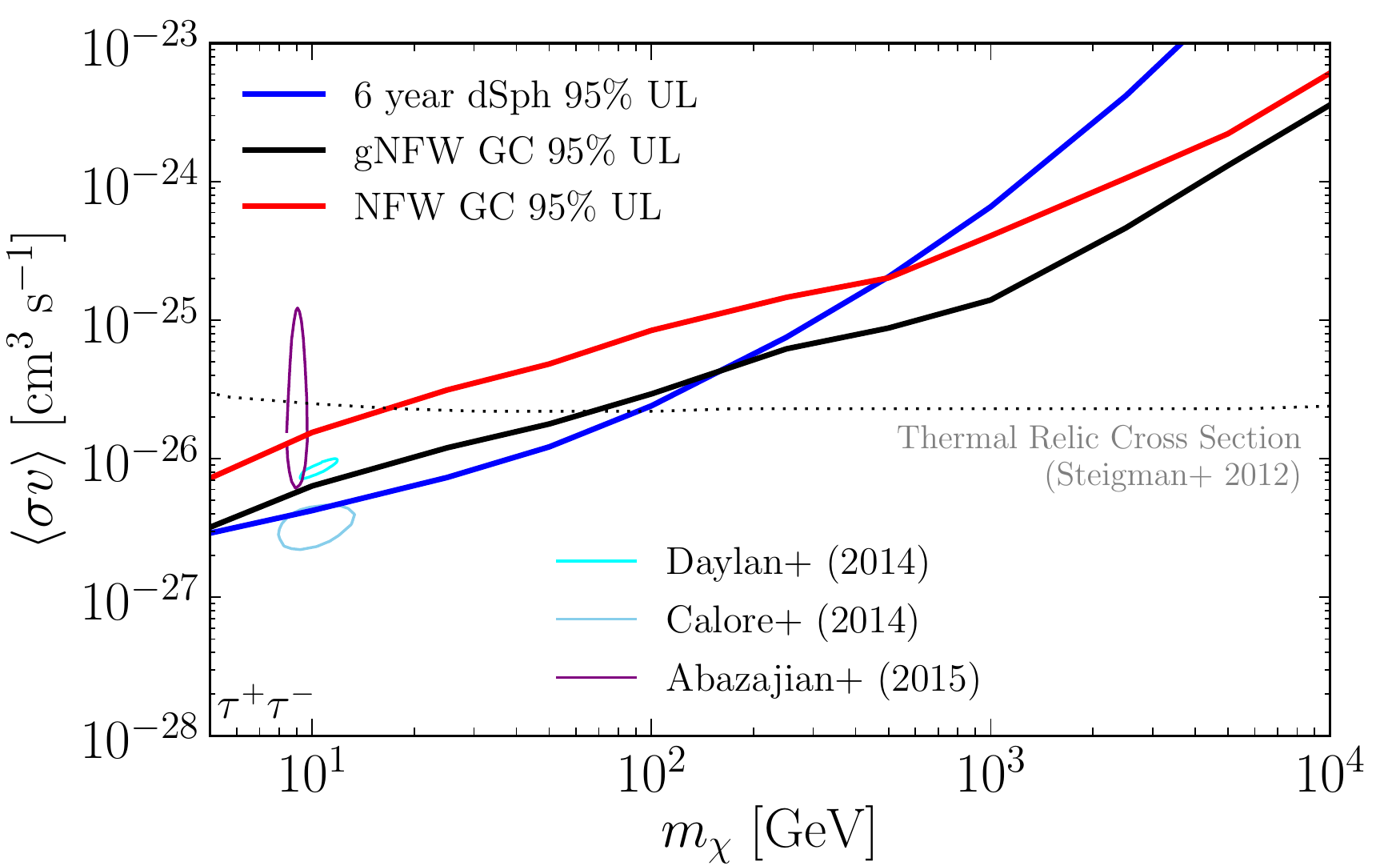}
\caption{Constraints on the dark matter annihilation cross section from Fermi's observations of the Galactic Center as a function of mass, for annihilations to $b\bar{b}$ (left) and $\tau^+ \tau^-$ (right) final states. Results are shown for the case of an NFW profile ($\gamma=1$) or a generalized NFW profile with $\gamma=1.25$ (in each case with $R_s=20$ kpc and a local density of 0.4 GeV/cm$^3$). These results are compared to the constraints derived from the stacked observations of Milky Way dwarf galaxies. From Ref.~\cite{TheFermi-LAT:2017vmf}.}
\label{limits}
\end{figure*}

Fermi's observations of the Galactic Center have been used to place some of the most stringent constraints on the dark matter annihilation cross section, and in Fig.~\ref{limits} these results are shown~\cite{TheFermi-LAT:2017vmf} (see also Ref.~\cite{Hooper:2012sr}). Results are presented for annihilations to $b\bar{b}$ and $\tau^+ \tau^-$ final states and for the case of an NFW profile ($\gamma=1$) or a generalized NFW profile with $\gamma=1.25$ (in each case with $R_s=20$ kpc and a local density of 0.4 GeV/cm$^3$). These constraints are compared to those derived from stacked observations of Milky Way dwarf galaxies. In evaluating such results, it is important to keep in mind that the constraints based on the Galactic Center can vary considerably depending on the assumptions made regarding the Milky Way's halo profile ({\it i.e.}\,the values of $\gamma$, $R_s$, $\rho_{\rm local}$).

\begin{center}
{\it 1. The Galactic Center Gamma-Ray Excess}
\end{center}

\begin{figure*}
\centering
\vspace{-1.5cm}
\vspace{0.6cm}
\hspace{0.6cm}
\includegraphics[width=0.75\textwidth,clip=true]{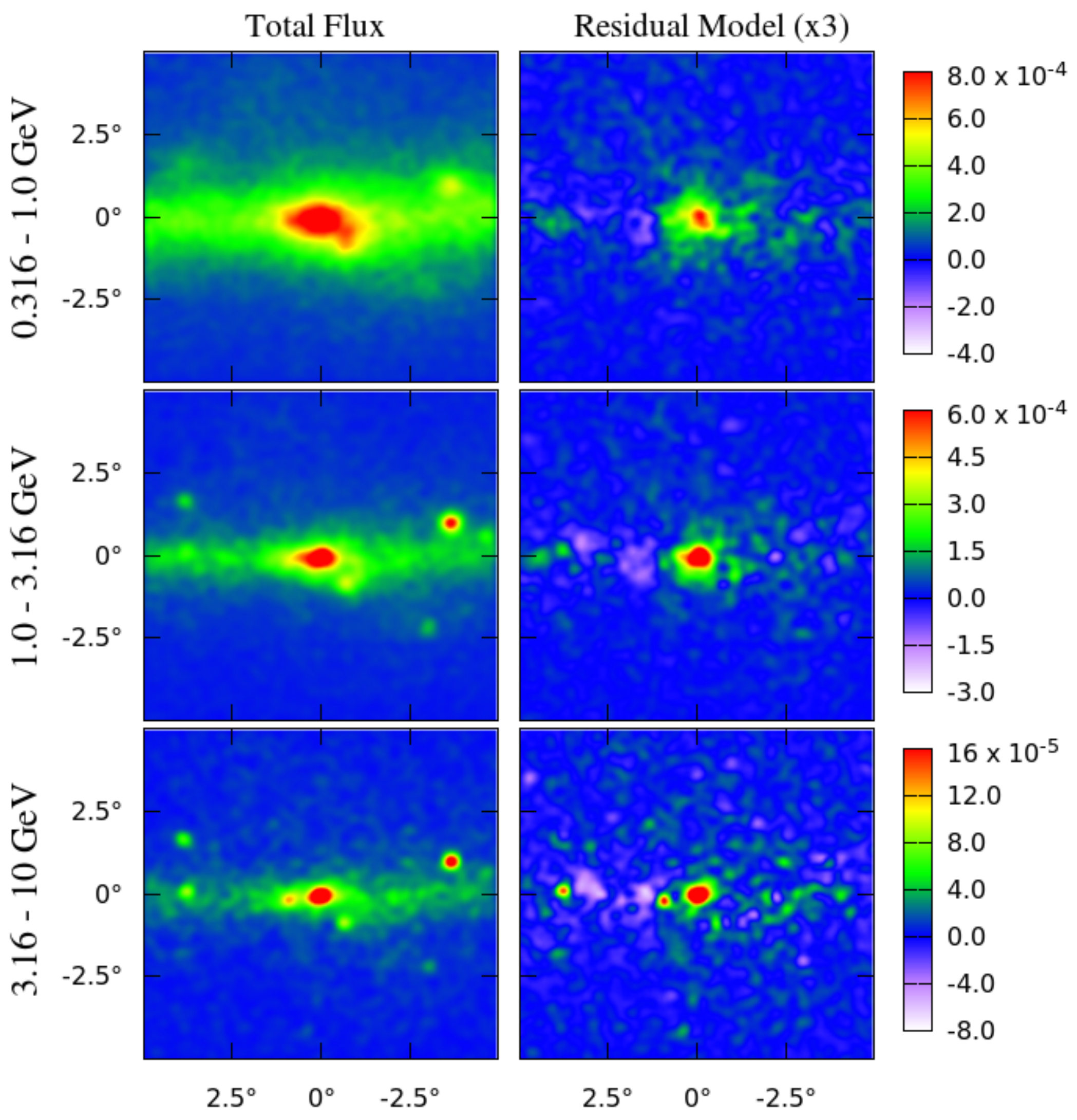}
\caption{The raw (left) and residual (right) intensity maps of the gamma-ray emission from the Inner Galaxy, as presented in Ref.~\cite{Daylan:2014rsa}. Although the existence of this excess was controversial for several years, by 2014 a consensus had begun to form that this signal is indeed present in the Fermi data. The origin of this emission remains hotly debated today.}
\label{radspec}
\end{figure*}

\begin{figure*}
\centering
\includegraphics[width=0.32\textwidth,clip=true]{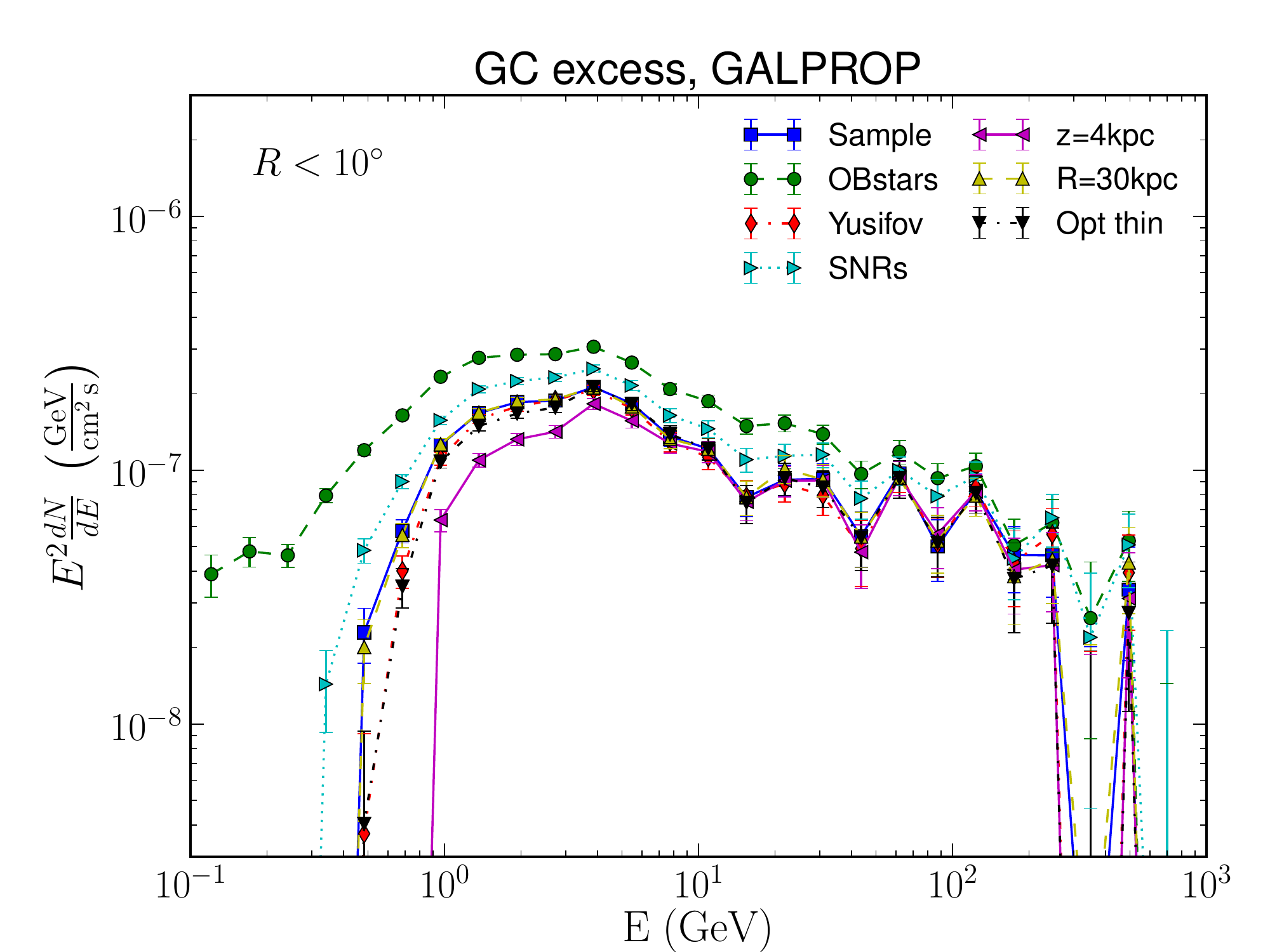} 
\includegraphics[width=0.32\textwidth,clip=true]{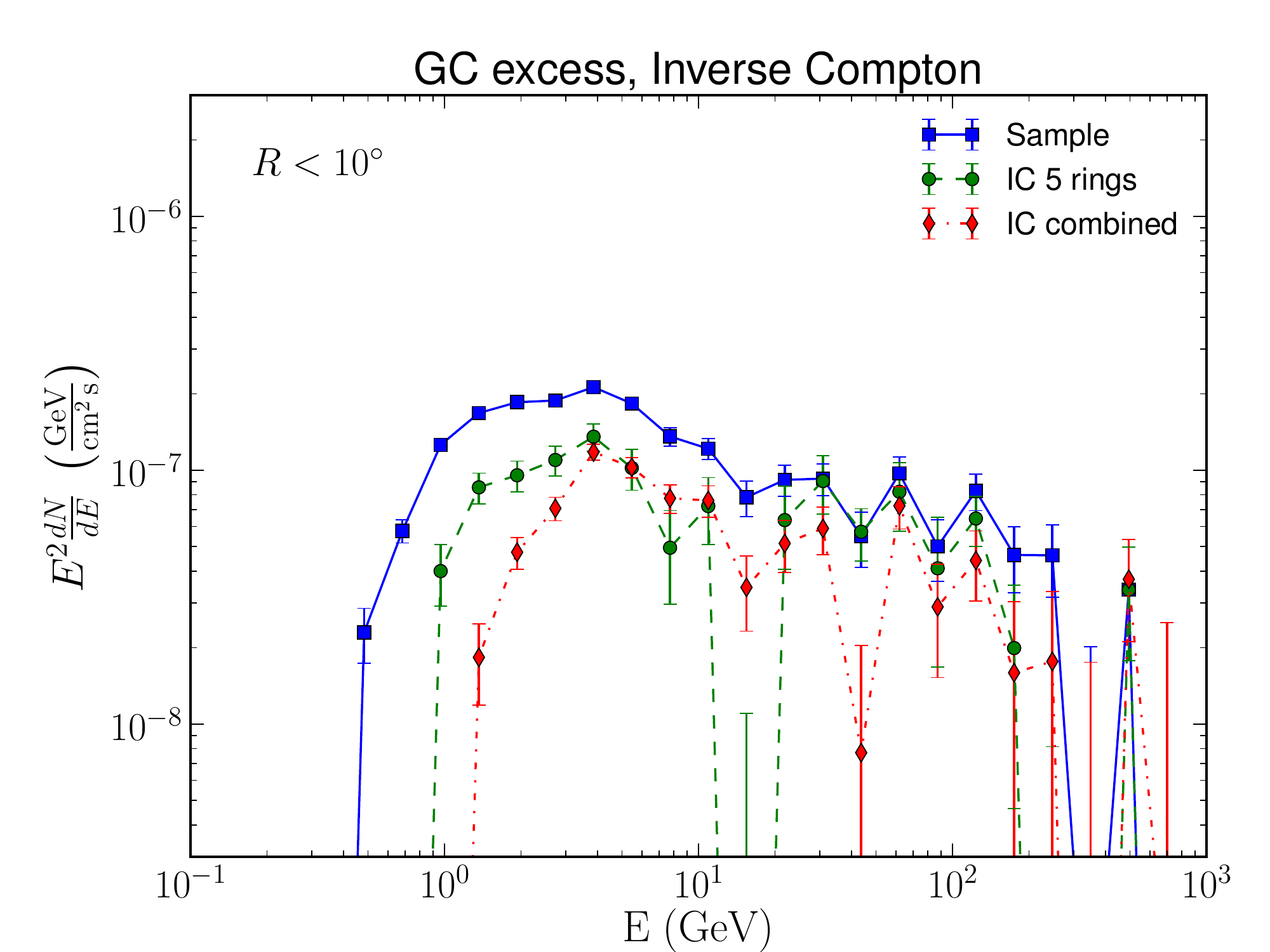}
\includegraphics[width=0.32\textwidth,clip=true]{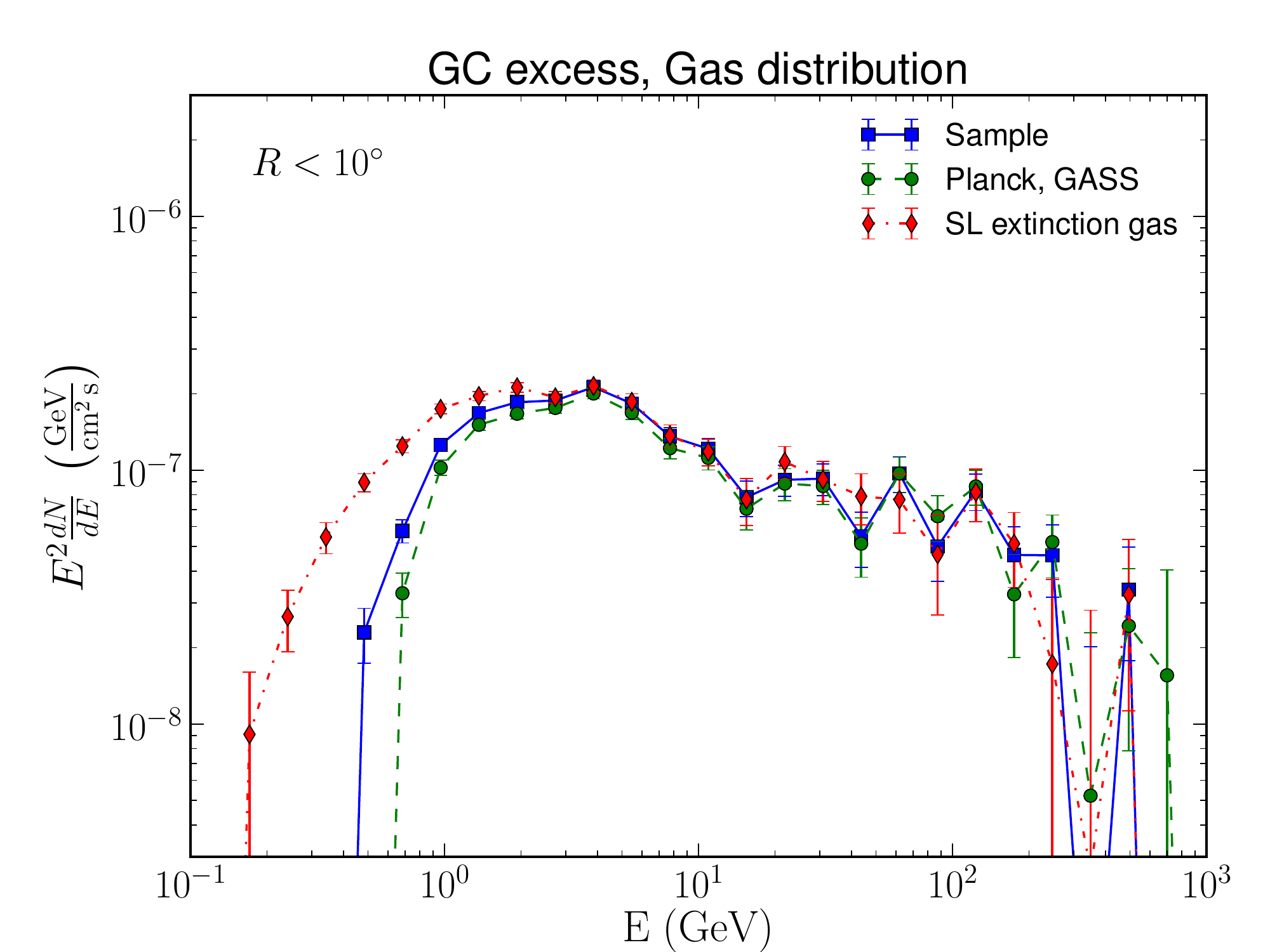} 
\includegraphics[width=0.32\textwidth,clip=true]{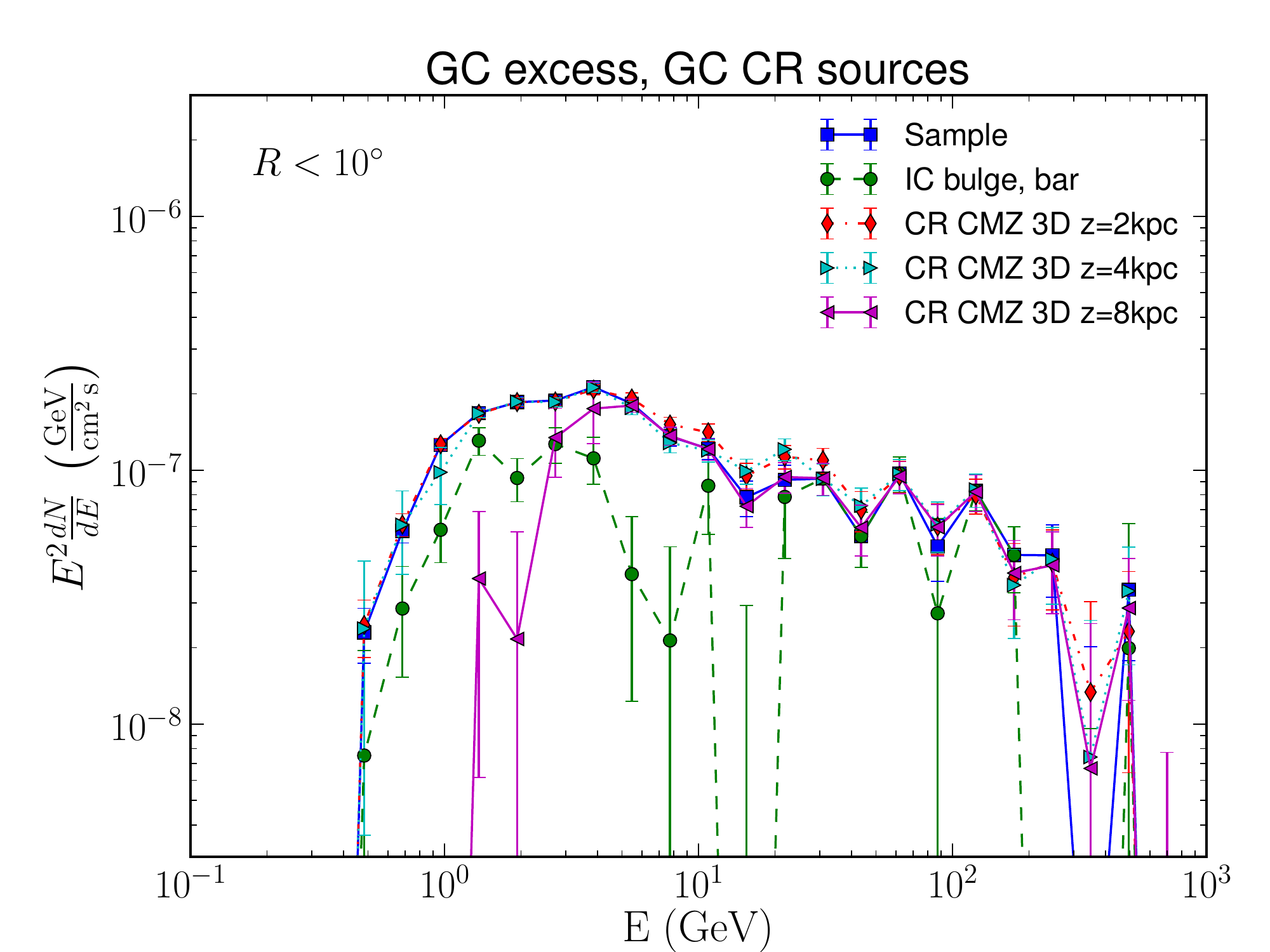}
\includegraphics[width=0.32\textwidth,clip=true]{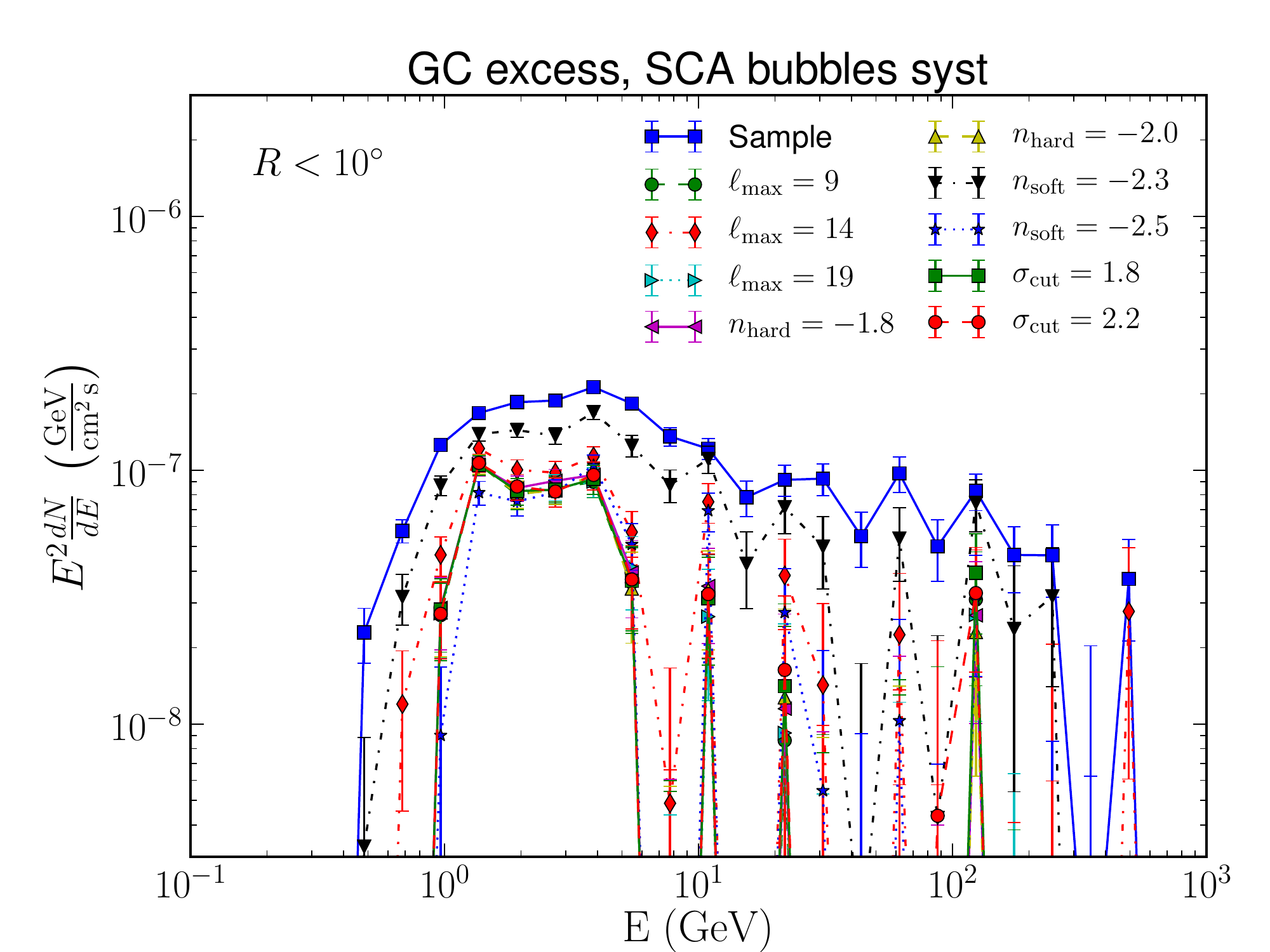}
\includegraphics[width=0.32\textwidth,clip=true]{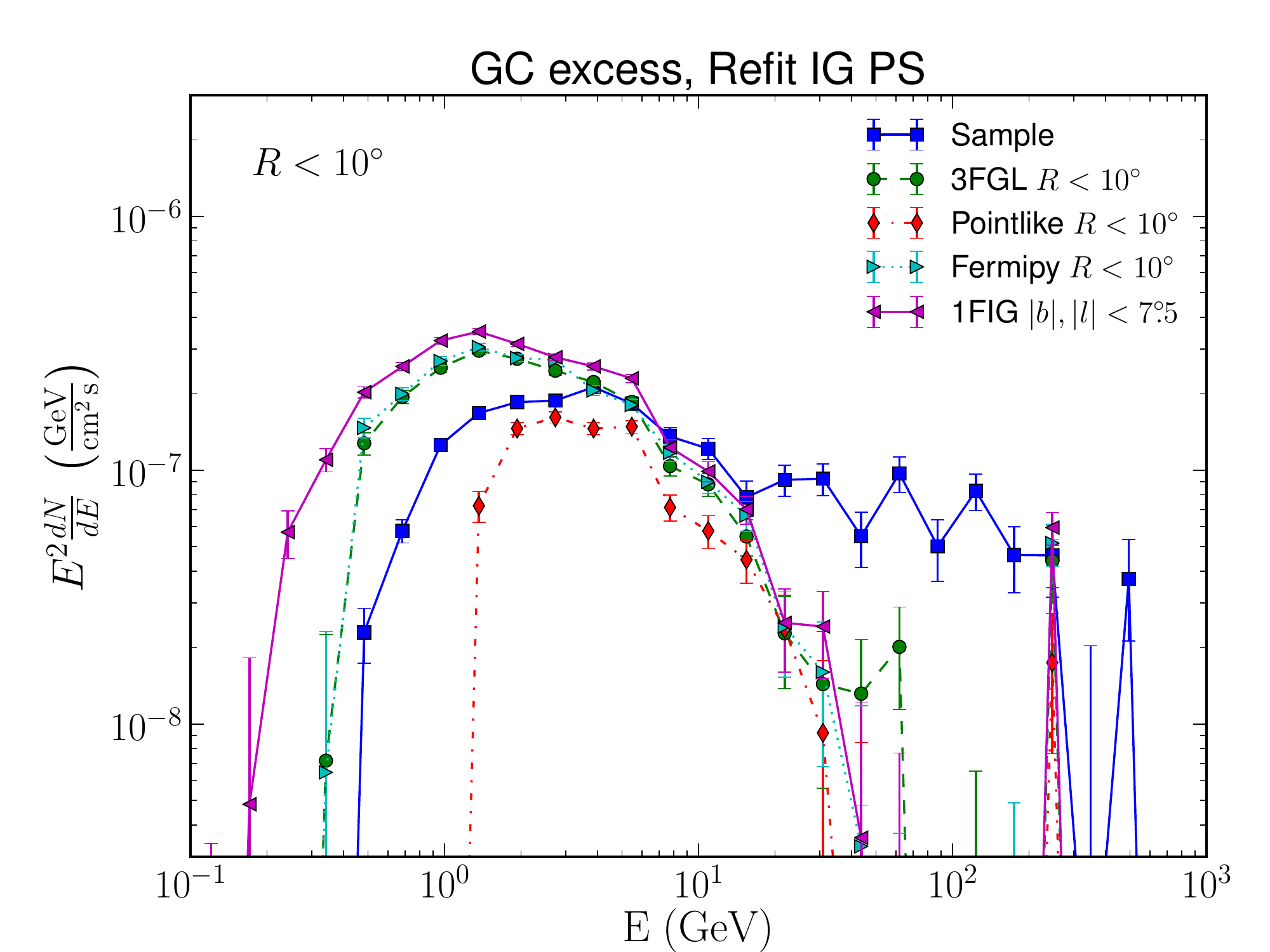}
\caption{The spectrum of the Galactic Center gamma-ray excess as presented in 2017 by the Fermi Collaboration, across a range of background models~\cite{TheFermi-LAT:2017vmf}. Although the detailed spectral shape of this signal can vary significantly depending on the background model that is adopted, the excess persists across all such variations.}
\label{spectrum}
\end{figure*}

\begin{figure*}
\centering
\includegraphics[width=0.6\textwidth,clip=true]{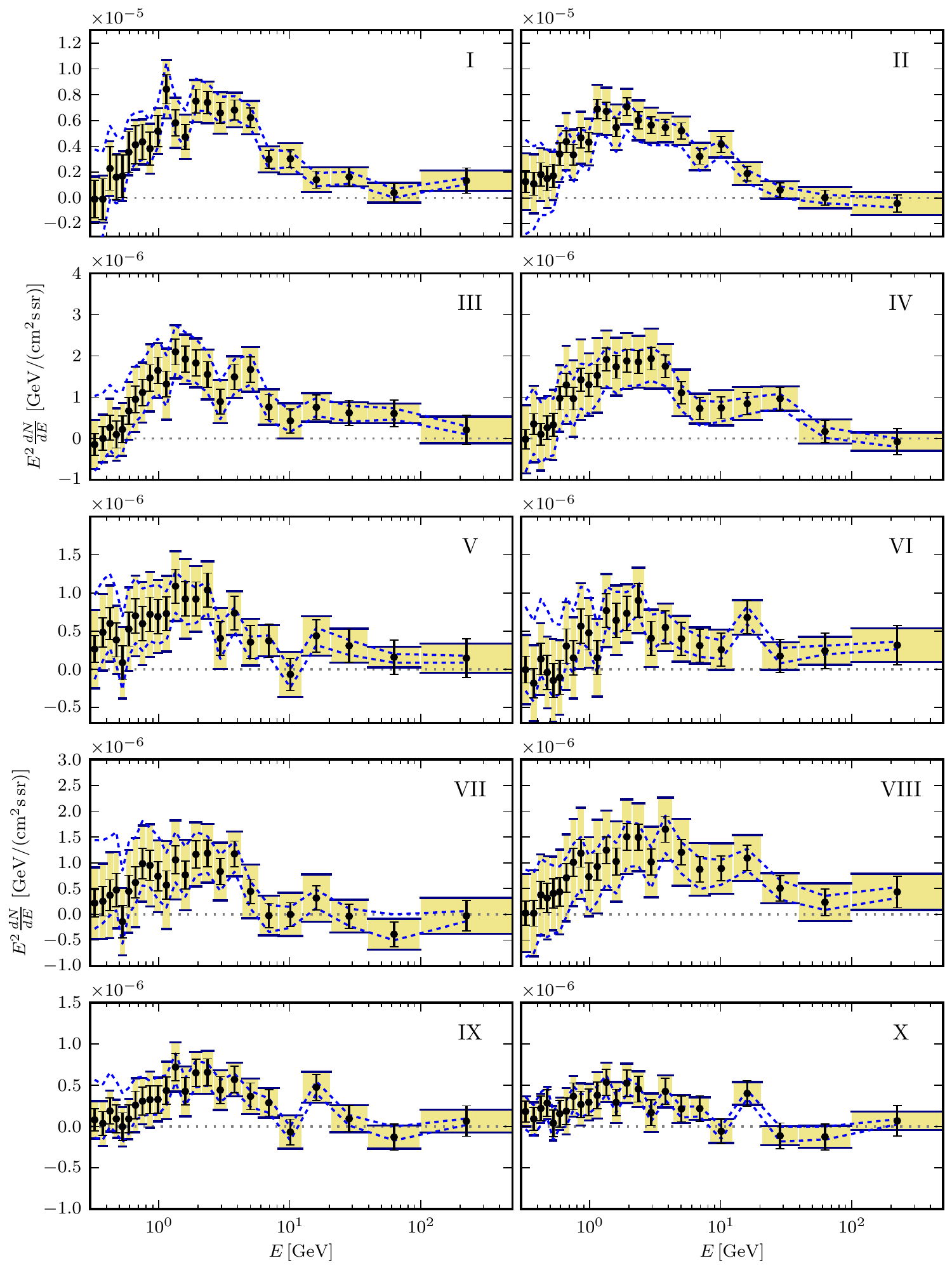}
\includegraphics[width=0.3\textwidth,clip=true]{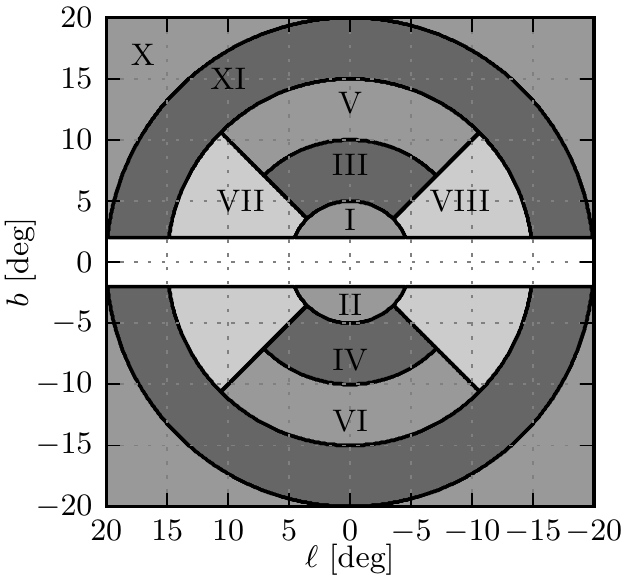} \\
\includegraphics[width=0.35\textwidth,clip=true]{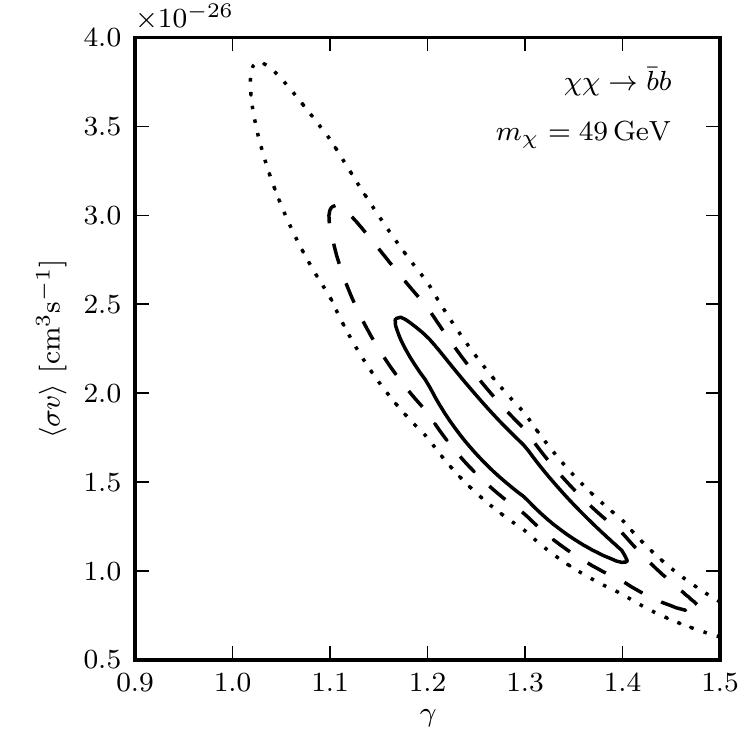} 
\includegraphics[width=0.35\textwidth,clip=true]{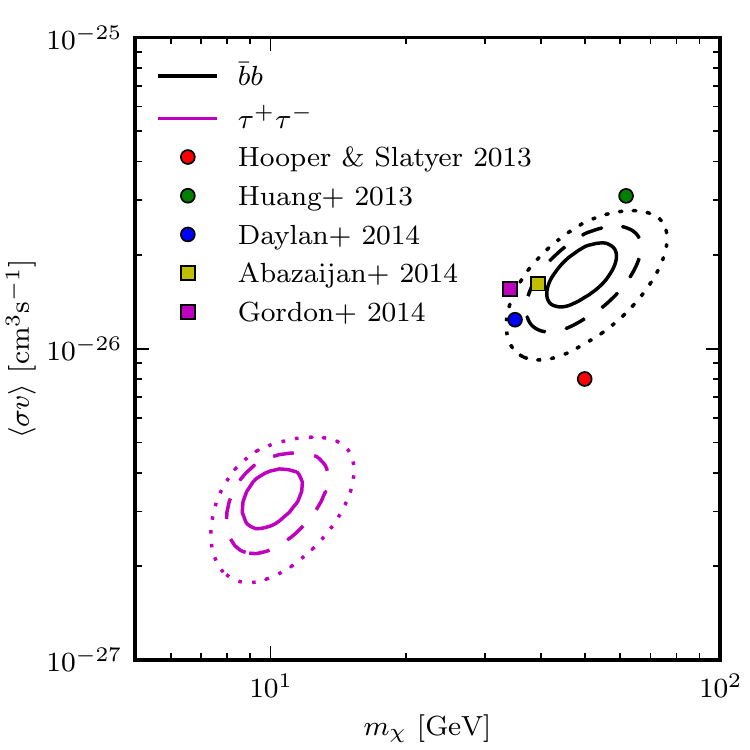}
\caption{Upper frames: The spectrum of the gamma-ray excess as measured in ten different sub-regions of the Inner Galaxy, as shown in the upper right frame. The spectral shape of this emission is approximately uniform across the region observed. Lower frames: The dark matter annihilation cross section, halo profile inner slope and mass required to accommodate the observed features of this excess. From Ref.~\cite{Calore:2014xka}.}
\label{calore2}
\end{figure*}

In 2009, Lisa Goodenough and I began to analyze the publicly available Fermi data in an effort to place constraints on any contribution from annihilating dark matter. In October of that year, we posted to the arXiv the first paper to identify what would become known as the Galactic Center gamma-ray excess~\cite{Goodenough:2009gk}. Over the following years, a number of studies~\cite{Hooper:2010mq,Hooper:2011ti,Abazajian:2012pn,Gordon:2013vta,Hooper:2013rwa,Huang:2013pda} improved upon this early work. By 2014 or so~\cite{Daylan:2014rsa}, a consensus had begun to form that the excess is in fact present, and exhibited the following characteristics:
\begin{itemize}
\item{The spectrum of the excess peaks at an energy of $\sim$\,1-5 GeV and falls off at both higher and lower energies (in $E^2 dN/dE$ units). The spectrum also appears to be uniform, without detectable variations throughout the Inner Galaxy~\cite{Calore:2014xka}. If interpreted as dark matter annihilation products, the spectral shape implies a dark matter candidate with a mass in the range of $\sim$\,40-70 GeV (for the case of annihilations to $b\bar{b}$). See Figs.~\ref{spectrum} and~\ref{calore2}.}
\item{The angular distribution of the excess is approximately azimuthally symmetric with respect to the Galactic Center, with a flux that scales as $F_{\gamma} \propto r^{-\Gamma}$ with $\Gamma=2.0-2.7$, where $r$ is the distance to the Galactic Center~\cite{Daylan:2014rsa,Abazajian:2014fta,Calore:2014xka,TheFermi-LAT:2015kwa,Linden:2016rcf,TheFermi-LAT:2017vmf}; see Fig.~\ref{calore2}. The emission continues with roughly this profile out to at least $10^{\circ}-20^{\circ}$ away from the Galactic Center (where it becomes too faint to reliably characterize). If interpreted in terms of dark matter annihilation, the observed morphology implies a halo profile with an inner slope of $\gamma = \Gamma/2 \sim 1.0-1.35$ (see Eq.~\ref{gennfw}).}
\item{The overall intensity of the excess is consistent with that expected from a dark matter candidate with an annihilation cross section of roughly $\langle \sigma v \rangle \sim 10^{-26}$ cm$^3/$s. See Fig.~\ref{calore2}.}
\end{itemize}

From early on, it was appreciated that these characteristics are each broadly consistent with the expectations of dark matter in the form of a simple annihilating relic. It was also realized, however, that the astrophysical backgrounds from this region were not particularly well understood. Of particular concern were those potential backgrounds associated with gamma-ray pulsars~\cite{Hooper:2010mq,Abazajian:2010zy,Hooper:2010mq,Abazajian:2012pn,Hooper:2013nhl,Gordon:2013vta,Yuan:2014rca,Abazajian:2014fta,Cholis:2014lta} and recent cosmic-ray outburst events~\cite{Carlson:2014cwa,Petrovic:2014uda,Cholis:2015dea}.

The gamma-ray emission observed from pulsars exhibits a spectral shape that is, in most cases, similar to that of the observed excess. Motivated by the possibility that the Galactic Center gamma-ray excess might originate from a population of unresolved gamma-ray pulsars, statistical tests were performed on the Fermi data to search for evidence of sub-threshold sources. In particular, Bartels, Krishnamurthy and Weniger utilized a wavelet-based technique designed to test for the presence of a large number of sub-threshold point sources~\cite{Bartels:2015aea}, while Lee, Lisanti, Safdi, Slatyer and Xue employed a non-Poissonian template fitting technique to a similar end~\cite{Lee:2015fea}. Each of these groups reported the detection of small-scale power in the photon distribution in the Inner Galaxy, and interpreted these results as evidence for a significant population of unresolved gamma-ray point sources. Today there is a consensus that the Fermi data from this region of the sky does, in fact, exhibit significant small scale power, possibly indicative of such a population. In my opinion, however, it is not at all clear that pulsars are responsible for the observed excess. While the small scale power reported in Refs.~\cite{Bartels:2015aea,Lee:2015fea} might reflect a large population of unresolved point sources, it is also entirely plausible that such a feature could arise as the result of imperfect modeling of diffuse backgrounds. Furthermore, if pulsars are responsible for the excess, it is somewhat surprising that we have not yet detected any individual pulsars from this region of the Galaxy~\cite{Hooper:2016rap,Cholis:2014lta,Bartels:2017xba}, or observed more low-mass X-ray binaries~\cite{Haggard:2017lyq}. In any case, it is clear that more data will be required to clarify this situation. Particularly promising are further gamma-ray observations of dwarf galaxies, as well as future radio searches for millisecond pulsars in the Inner Galaxy~\cite{Calore:2015bsx}.

\subsection{The Isotropic Gamma-Ray Background}

Dark matter annihilations could produce significant contributions to the isotropic gamma-ray background (IGRB). In particular, the IGRB is expected to receive contributions from dark matter annihilating in the halo of the Milky Way, as well as from the integrated annihilation rate over the large scale structure of the universe. Furthermore, over cosmological distances, a significant fraction of high-energy gamma rays scatter with the cosmic radiation backgrounds, producing $e^+ e^-$ pairs which then go on to generate additional gamma rays as part of an electromagnetic cascade. 

The Fermi Collaboration has measured the IGRB at energies between 100 MeV to 820 GeV~\cite{Ackermann:2014usa}.  Although previously detected by other instruments~\cite{1978ApJ...222..833F,Sreekumar:1997un}, Fermi's measurement of the IGRB has provided a more detailed description of its characteristics and led to a more complete understanding of its origin.

It has long been speculated that the majority of the IGRB is produced by a large number of unresolved sources, such as active galactic nuclei (AGN)~\cite{Stecker:1993ni,1993MNRAS.260L..21P, Salamon:1994ku,Stecker:1996ma,Mukherjee:1999it,Narumoto:2006qg,Giommi:2005bp,Dermer:2006pd,Pavlidou:2007dv,Inoue:2008pk} and star-forming galaxies~\cite{Pavlidou:2002va,Thompson:2006qd,Fields:2010bw,Makiya:2010zt}, perhaps along with contributions from the annihilations or decays of dark matter particles~\cite{Stecker:1978du,Gunn:1978gr,Gao:1991rz,Ullio:2002pj}. In recent years, Fermi's detection of gamma-ray emission from both non-blazar AGN~\cite{Inoue:2011bm} and star-forming galaxies~\cite{Ackermann:2012vca}, combined with the observed correlations of the emission at gamma-ray and radio/infrared wavelengths, has revealed that these source classes each contribute significantly to the IGRB. Even more recent studies have shown that the combination of these source classes dominates the observed IGRB~\cite{hooper2016radio,Linden:2016fdd}, leaving relatively little room for the presence of dark matter annihilation products.

The high-latitude gamma-ray sky receives contributions from several different processes associated with the annihilation of dark matter particles. First, dark matter particles annihilating in the halo of the Milky Way generate a flux of gamma rays that can be calculated using Eq.~\ref{gamma}. While this emission is not strictly isotropic, the line-of-sight integral in Eq.~\ref{gamma} departs by less than 10\% from the average value within the range of angles that contribute to Fermi's measurement of the IGRB ($|b| > 20^{\circ}$). A component of gamma-ray emission with such a small degree of variation across the high-latitude sky would be indistinguishable from the overall IGRB.

Dark matter annihilations beyond the boundaries of the Milky Way also contribute to the IRGB. Over cosmological distances, however, gamma rays are much more likely to be attenuated via pair production, thereby initiating electromagnetic cascades. Neglecting attenuation for the moment, the spectrum of gamma rays per area per time per solid angle from annihilating dark matter is given by:
\begin{equation}
\frac{dN_{\gamma}}{dE_{\gamma}}(E_{\gamma}) = \frac{c}{8\pi} \int \frac{\langle \sigma v \rangle \rho^2_{\rm X}(z) \, dz}{H(z) (1+z)^3 \, m^2_{\rm X}}  \,\bigg(\frac{dN_{\gamma}}{dE'}\bigg)_{E' = E_{\gamma}(1+z)},  
\end{equation}
where $H(z) = H_0 [\Omega_M (1+z)^3 + \Omega_{\Lambda}]^{0.5}$ is the expansion rate of the universe in terms of the cosmological parameters $\Omega_M=0.31$, $\Omega_{\Lambda} =0.69$ and $H_0=67.7$ km/s~\cite{Aghanim:2018eyx}. Although the average dark matter density evolves as $\rho_{\rm X}(z) \propto (1+z)^3$, the clumping of dark matter into halos plays a very important role in this integral, effectively boosting the annihilation rate to potentially observable levels. Lastly, the quantity $dN_{\gamma}/dE'$ is the gamma-ray spectrum produced per annihilation, after accounting for the effects of cosmological redshift.

High-energy gamma rays are significantly attenuated through their scattering with infrared, optical and microwave radiation fields~\cite{Murase:2011yw,Murase:2012xs,Murase:2011cy,Murase:2012df,Berezinsky:2016feh}. The inverse mean free path of these interactions is given by:
\begin{eqnarray}
l^{-1} (E_{\gamma},z)&=&  \int  \sigma_{\gamma\gamma}(E_{\gamma},\epsilon) \, \frac{dn}{d\epsilon}(\epsilon,z) \, d\epsilon, 
\end{eqnarray}
where $\sigma_{\gamma\gamma}$ is the total pair-production cross section~\cite{aharonian1983} and $dn(\epsilon,z)/d\epsilon$ is the differential number density of target photons at redshift, $z$~\cite{Dominguez:2010bv}. In practice, such interactions make the universe opaque to photons with energies greater than a few hundred GeV, causing the photons and electrons above this threshold to have their energy transferred into an electromagnetic cascade with a universal spectrum that peaks mildly at $\sim$\,100 GeV and extends across the entire range of energies measured by Fermi. 

Fermi's measurement of the IGRB has been used to produce constraints on the dark matter's annihilation cross section that are competitive with, although slightly weaker than, those derived from observations of dwarf galaxies and the Galactic Center~\cite{Ackermann:2015tah,DiMauro:2015tfa,Ajello:2015mfa,Cholis:2013ena,Ando:2015qda}. Given that these constraints rely on a different set of astrophysical uncertainties, they remain relevant through their complementarity to these other techniques.

\section{Cosmic-Ray Searches for Dark Matter Annihilation Products}

The cosmic-ray spectrum is dominated by protons and nuclei. At energies below the ``knee'' ($E_{\rm knee} \sim 10^{6}$ GeV), most of these particles are thought to originate from Galactic supernova remnants. At higher energies, this spectrum is instead dominated by particles that originate from sources beyond the boundaries of the Milky Way, perhaps including active galactic nuclei. At all energies, the cosmic-ray spectrum is dominated by matter over antimatter. It has long been appreciated that if dark matter particles are annihilating or decaying in the halo of the Milky Way, such processes would (in most models) produce equal amounts of matter and antimatter, leading to an excess of antimatter relative to that predicted by standard astrophysical mechanisms. In this sense, indirect searches for dark matter using cosmic rays are often (but not always) searches for cosmic-ray antimatter, such as antiprotons, positrons, or anti-deuterons~\cite{Silk:1984zy,Ellis:1988qp,Stecker:1985jc,Turner:1989kg,Kamionkowski:1990ty,Bergstrom:1999jc,Donato:2003xg,Bringmann:2006im,Donato:2008jk,Fornengo:2013xda,Hooper:2014ysa,Pettorino:2014sua,Boudaud:2014qra,Cembranos:2014wza,Cirelli:2014lwa,Bringmann:2014lpa,Giesen:2015ufa,Evoli:2015vaa}.

Once cosmic rays are injected into the halo, they move via diffusion, random walking through the Milky Way's tangled magnetic field, while also undergoing interactions that can lead to energy losses, decay, etc. These processes are collectively described by the cosmic-ray transport equation~\cite{Strong:2007nh}:
\begin{eqnarray}
\frac{\partial{}}{\partial{t}}\frac{dn}{dE}(E,\vec{x},t) =  \vec{\bigtriangledown} \cdot \bigg[D(E,\vec{x}) \vec{\bigtriangledown} \frac{dn}{dE}(E,\vec{x},t) \bigg] + \frac{\partial}{\partial E} \bigg[\frac{dE}{dt}(E) \, \frac{dn}{dE}(E,\vec{x},t)    \bigg]  + Q(E,\vec{x},t), \nonumber \\
\label{diffusionlosseq}
\end{eqnarray}
where $dn/dE$ is the differential number density of cosmic rays, $D$ is the diffusion coefficient, and the source term, $Q$, describes the spectrum, distribution, and time profile of cosmic rays injected into the halo (or removed from the halo in case of decay or spallation). This equation is generally solved in the steady-state limit (setting the left side equal to zero), and for a set of boundary condition. More specifically, a cylindrical geometry is generally adopted, enclosing a volume with a half-thickness of $L_{\rm z} \sim (1-6) \, {\rm kpc}$ and a radius of $\sim$\,20 kpc, beyond which the particles are not confined by the Galactic Magnetic Field and freely escape. 

The source term in Eq.~\ref{diffusionlosseq} includes contributions from individual sources of cosmic-ray electrons (supernova remnants, pulsars, etc.), as well as secondary particles, which are produced through the interactions of other cosmic rays. Secondary electrons and positrons, for example, are generated in the decays of pions and kaons that are produced in the collisions of hadronic cosmic rays with gas. The flux of cosmic-ray secondaries can be calculated from Eq.~\ref{diffusionlosseq} by setting $Q=\int J_p n_{\rm gas}(d\sigma/dE) dE_p$, where $J_p$ is the flux of hadronic cosmic rays, $n_{\rm gas}$ is the gas density, and $d\sigma/dE$ is the differential cross section for the production of secondaries~\cite{Moskalenko:1997gh,Moskalenko:2001ya}.

Detailed models of cosmic-ray transport have many free parameters, including the spatial distribution and spectrum of sources, the energy and spatial dependance of the diffusion constant, the boundary conditions of the diffusion zone, as well as those which account for effects such as convection, diffusive reacceleration and solar modulation. Fortunately, there are also many independent observations that can be used to constrain these parameters. In particular, by measuring the energy-dependent ratios of secondary-to-primary cosmic rays, we can infer a great deal about cosmic-ray transport. In particular, stable secondary-to-primary ratios (such as boron-to-carton) inform us about the average amount of matter traversed by cosmic rays as a function of energy.  Unstable secondary-to-primary ratios, in contrast, serve as a measurement of the amount of time that cosmic rays have been propagating. Beryllium-10 is particularly useful in this regard, being the longest lived and best measured unstable secondary. The measurement of $^{10}$Be/$^{9}$Be thus serves as a clock, since the ratio of the radioactive isotope to the stable one is directly related to the amount of time elapsed since the creation of the particles. When global fits are performed to the current set of cosmic-ray data, one finds $D(E) \simeq (3.9 \times 10^{28} \, {\rm cm}^3/{\rm s})\,E^{0.3}$ and $L_z \simeq 4$ kpc.

To build some intuition for cosmic-ray transport, consider the simple example of a burst-like source of cosmic-ray protons that occurs at a particular place and time within the diffusion zone of the Galaxy. In this case, the source term is given by $Q = Q_0 \delta(\vec{x})\delta(t)$, and for protons we can safely neglect any energy losses. The transport equation then reduces to the following:
\begin{equation}
\frac{\partial{}}{\partial{t}}n(\vec{x},t) =  \vec{\bigtriangledown} \cdot \bigg[D(\vec{x}) \vec{\bigtriangledown} n(\vec{x},t) \bigg]  + Q_0 \delta(\vec{x})\delta(t),
\end{equation}
whose solution is $n \propto (Dt)^{-3/2} \exp(-r^2/4Dt)$, where $r$ is the radial distance to the source. The main feature of this solution is that these particles diffuse outward from the source a distance of order $L_{\rm dif} \sim \sqrt{D t}$, which scales with the square root of time as one would expect for a random walk. For a diffusion constant of $D = 3.9 \times 10^{28} \times (E/\rm GeV)^{0.3}$, the diffusion length is given by $L_{\rm dif} \sim 800 \, {\rm pc} \times (E/100 \, {\rm GeV})^{1/6} (t/{\rm Myr})^{1/2}$. Note that this result is easily generalizable to the case of an arbitrary injection history by summing the results for a series of burst-like events.

In the case of cosmic-ray electrons and positrons, it is important to include the effects of energy losses from inverse Compton scattering and synchrotron~\cite{Blumenthal:1970gc}:
\begin{eqnarray}
\label{losses}
-\frac{dE_e}{dt}(r) &=& \sum_i \frac{4}{3}\sigma_T \rho_i(r) S_i(E_e) \bigg(\frac{E_e}{m_e}\bigg)^2 + \frac{4}{3}\sigma_T \rho_{\rm mag}(r) \bigg(\frac{E_e}{m_e}\bigg)^2  \\
& \approx & 1.02 \times 10^{-16} \, {\rm GeV}/{\rm s} \, \times \bigg[ \sum_i \bigg(\frac{\rho_{i}(r)}{{\rm eV}/{\rm cm}^3}\bigg) \, S_{i}(E_e) + 0.224 \,\bigg(\frac{B}{3\, \mu \rm{G}}\bigg)^2 \bigg]  \,  \bigg(\frac{E_e}{{\rm GeV}}\bigg)^2, \nonumber
\end{eqnarray}
where $\sigma_T$ is the Thomson cross section and the sum is carried out over the various components of the radiation backgrounds, such as the cosmic microwave background (CMB) and starlight, as well as infrared and ultraviolet emission. The quantity, $S$, is the Klein-Nishina factor, which suppresses inverse Compton scattering at very high energies ($E_e \gsim m^2_e/2T$)~\cite{Longair}:
\begin{equation}
S_i (E_e) \approx \frac{45 \, m^2_e/64 \pi^2 T^2_i}{(45 \, m^2_e/64 \pi^2 T^2_i)+(E^2_e/m^2_e)}.
\end{equation}

If we consider a burst-like source of cosmic-ray electrons/positrons, we find that energy losses limit the distance that such particles can propagate, especially at high energies. It follows from Eq.~\ref{losses} that a 100 GeV (1 TeV) electron will lose most of its energy over a timescale of $t_{\rm loss} \sim 3$ Myr (300 kyr), over which $L_{\rm dif} \sim 1.4 \, {\rm kpc}$ (400 pc). From this exercise, we conclude that only a relatively small volume of the local Galaxy contributes to the observed high-energy cosmic-ray electron/positron spectrum.

\subsection{Cosmic-Ray Positrons}

In 2008, the collaboration of scientists operating the satellite-based experiment PAMELA reported that the cosmic-ray positron fraction (the ratio of positrons to positrons-plus-electrons) rises between approximately 10 GeV and 100 GeV~\cite{Adriani:2008zr} (see also Ref.~\cite{Adriani:2013uda}). While consistent with previous indications from HEAT~\cite{Barwick:1997ig} and AMS-01~\cite{Aguilar:2007yf}, this rise is in stark contrast to the behavior expected for a positron spectrum dominated by secondary particles produced during cosmic-ray propagation~\cite{Moskalenko:1997gh}. Within this context, the possibility that annihilating dark matter might be responsible for this signal generated a great deal of interest~\cite{Cholis:2008hb,Bergstrom:2008gr,Zurek:2008qg,Harnik:2008uu,Cirelli:2008jk}, although it was also pointed out that nearby pulsars~\cite{Hooper:2008kg} or the acceleration of secondary positrons in supernova remnants~\cite{Blasi:2009hv,Blasi:2009bd} could potentially account for the excess positrons. In any case, the rising positron fraction requires a source (or sources) of cosmic-ray positrons beyond that associated with standard secondary production.

\begin{figure*}
\begin{centering}
\includegraphics[width=3.30in,angle=0]{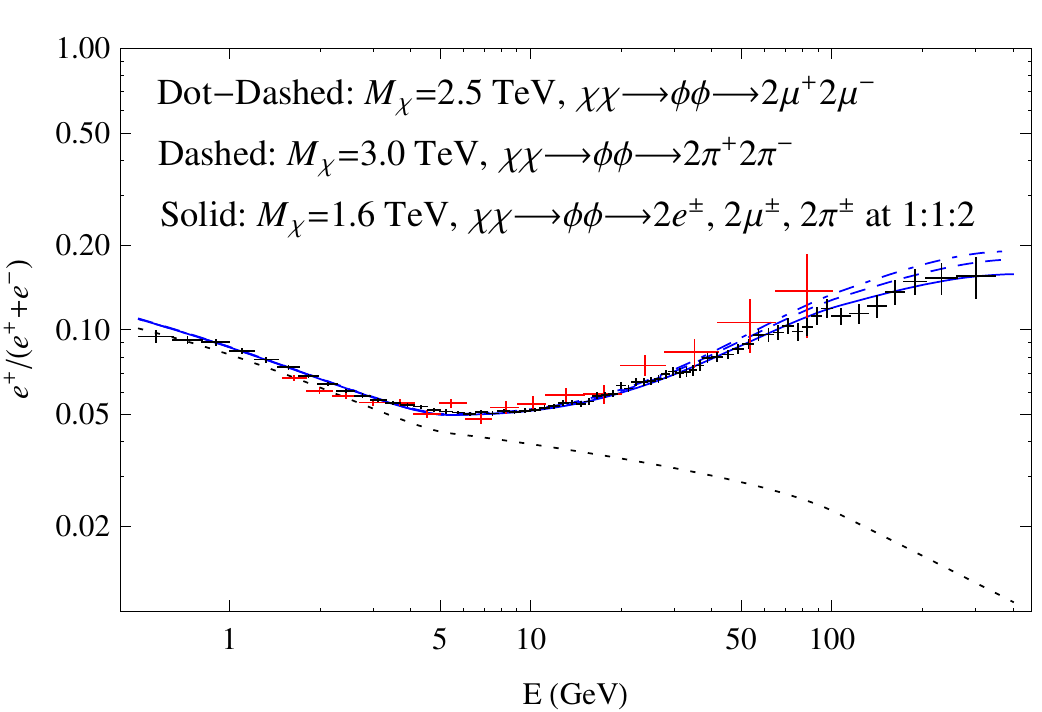}
\includegraphics[width=3.30in,angle=0]{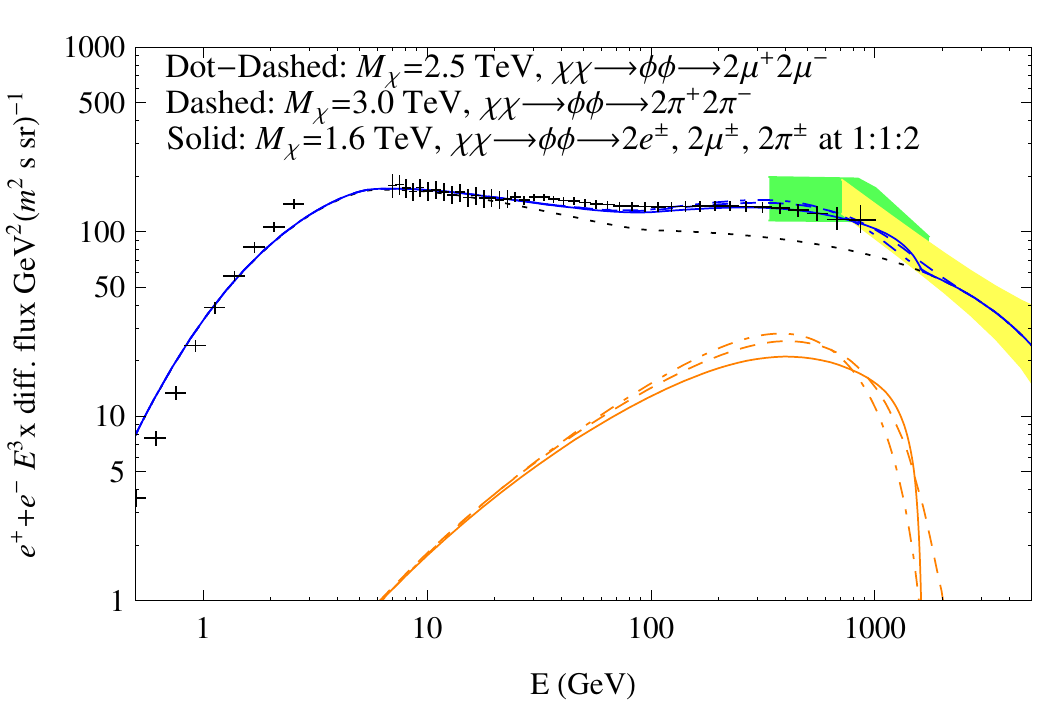} \\
\end{centering}
\caption{The cosmic-ray positron fraction (left) and electron+positron spectrum (right) in models in which the dark matter annihilates into a pair of intermediate states, $\phi$, which proceed to decay to to $\mu^+ \mu^-$, $\pi^+ \pi^-$, or to a 1:1:2 ratio of $e^+ e^-$, $\mu^+ \mu^-$, and $\pi^+ \pi^-$. The error bars shown represent the positron fraction as measured by AMS-02 (black, left) and PAMELA (red, left), and the electron+positron spectrum as measured by Fermi and AMS-01 (black, right). In each case, the parameters of the Galactic cosmic-ray transport model were selected in order to provide a good fit to the various secondary-to-primary ratios. The expected backgrounds from standard secondary production are shown as black dotted lines. From Ref.~\cite{Cholis:2013psa}.}
\label{fig:cholis}
\end{figure*}

When the data from PAMELA was combined with that from AMS-02, as well as the electron+positron spectrum from Fermi, it became clear that in order for dark matter to generate this signal, the particle would have to be quite heavy ($\sim$\,1-3 TeV) and annihilate into light intermediate states~\cite{ArkaniHamed:2008qn,Cholis:2008qq}; see Fig.~\ref{fig:cholis}~\cite{Cholis:2013psa}. Light mediators could also induce Sommerfeld enhancements, thereby allowing a heavy thermal relic to generate a large observed flux of positrons.

If astrophysical sources were responsible for the rising positron fraction, those sources must reside within several hundred parsecs of the Solar System, due to the rapid energy losses of high-energy electrons/positrons. Within this context, the nearby pulsars Geminga and Monogem (also known as B0656+14) are particularly interesting. In fact, it was shown in Ref.~\cite{Hooper:2008kg} that if these sources deposited on the order of 10\% of their energy budget into high-energy electron-positron pairs, they could account for the observed positron excess. 

In 2017, the HAWC Collaboration released their first measurements of the very high-energy (multi-TeV) gamma-ray emission from the Geminga and Monogem pulsars~\cite{Abeysekara:2017hyn}, finding that the emission from these sources follows a diffusive profile extending out to at least $\sim$\,5$^\circ$ in radius (corresponding to a physical extent of $\sim$\,25~pc)~\cite{Abeysekara:2017old}. The spatially extended nature of this emission indicates that it is generated through the inverse Compton scattering of very high-energy electrons and positrons with the cosmic microwave background and other radiation fields. Furthermore, the fluxes of very high-energy gamma-rays observed from Geminga and Monogem indicate that these sources inject a flux of positrons into the local interstellar medium that is approximately equal to the value required to account for the observed positron excess. This new information strongly favors the conclusion that the positron excess is generated by nearby pulsars, diminishing the motivation for annihilating dark matter or other exotic mechanisms~\cite{Hooper:2017gtd}.

\begin{figure}[t!]
  \includegraphics[width=0.49\linewidth]{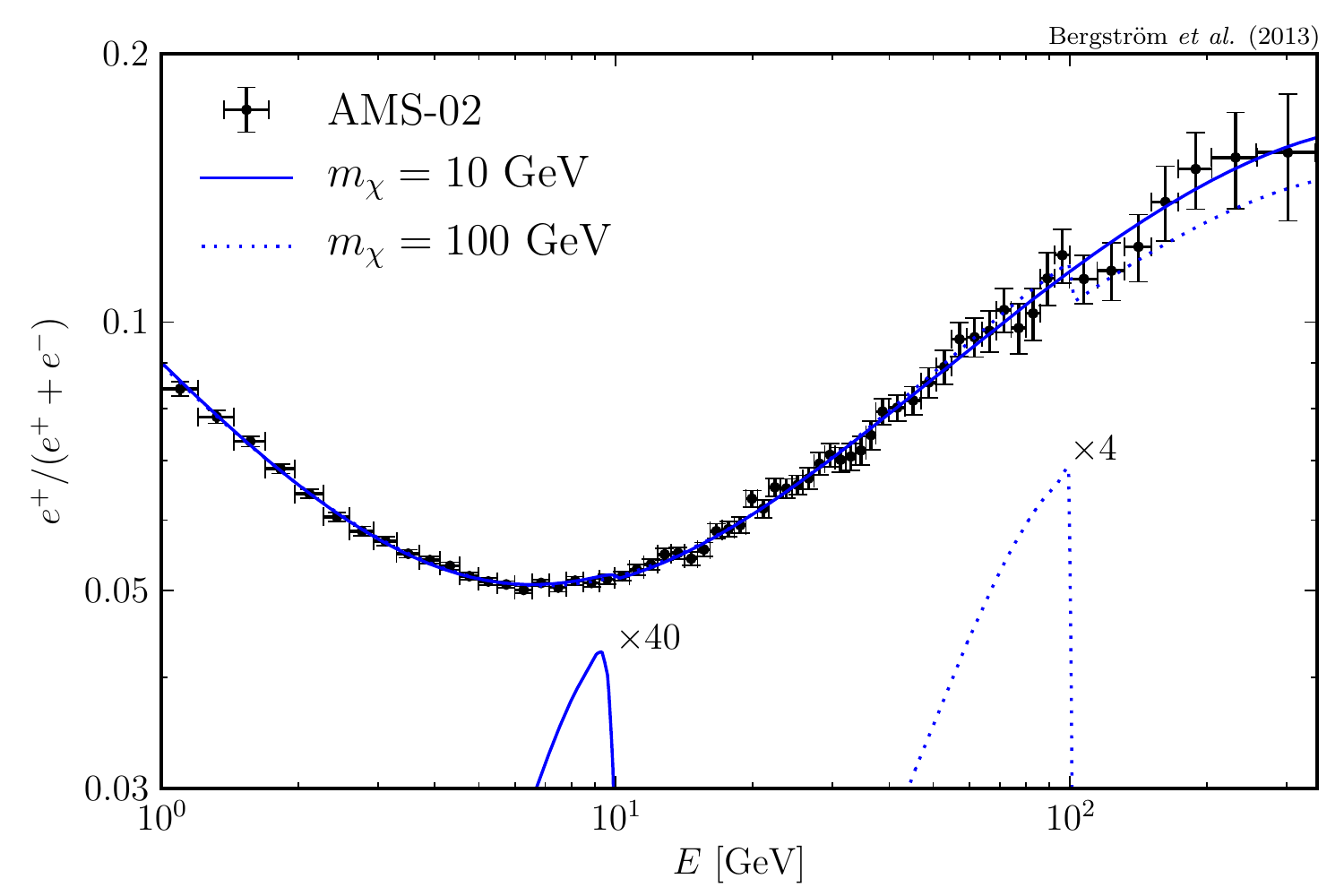}
      \includegraphics[width=0.49\linewidth]{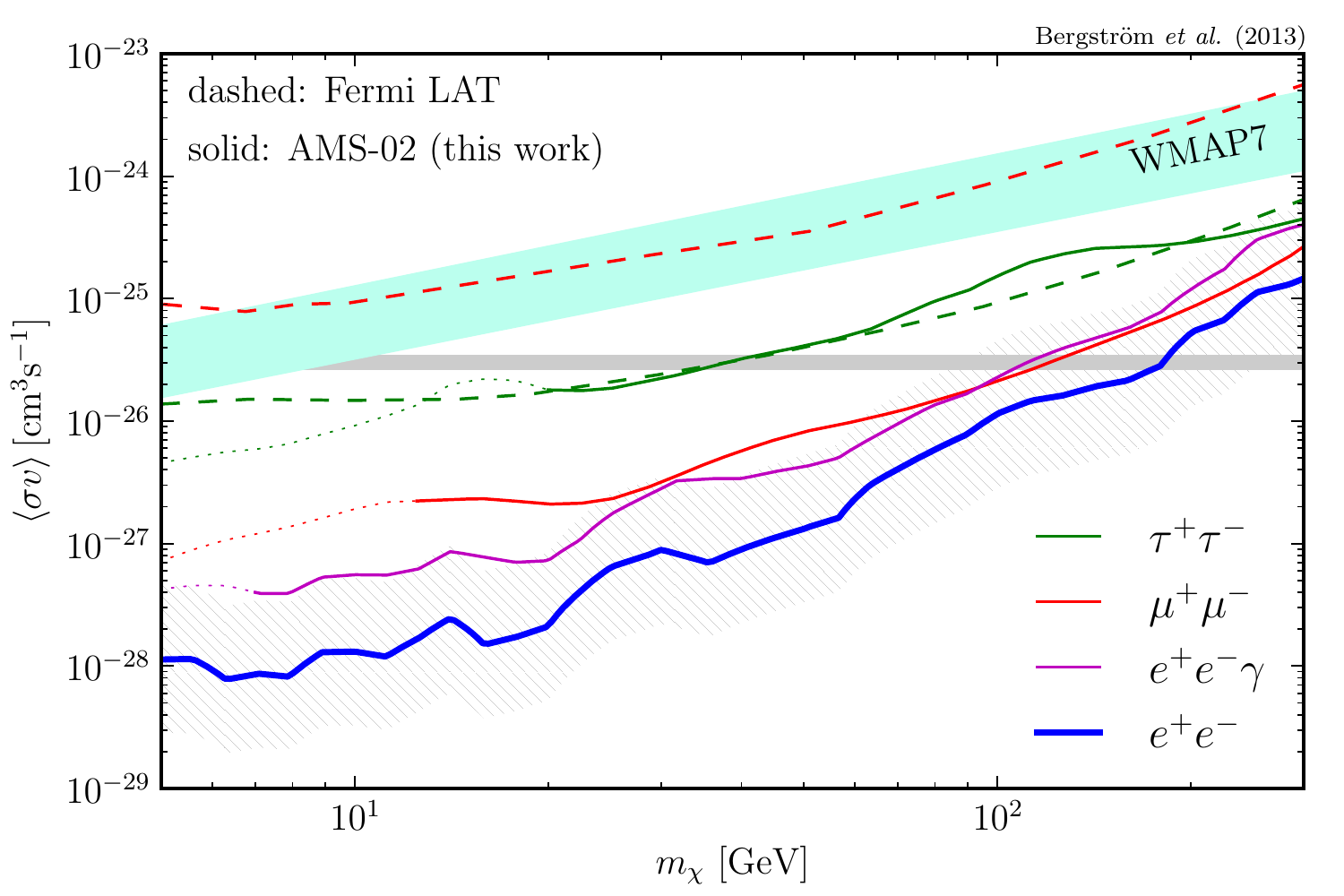}
  \caption{Left: The AMS positron fraction as measured by AMS-02 and 
  background+signal fit for dark matter annihilating directly to $e^+ e^-$, for dark matter masses of 10 or 100 GeV. The normalization of the dark matter signal in each case was chosen such that it is excluded
    at the $95\%$ confidence level. For visibility, the contribution from dark matter
    (lower lines) has been rescaled as indicated. Right: Upper limits ($95\%$ confidence level) on the 
  dark matter annihilation cross section, as derived from the
    AMS-02 positron fraction, for various leptonic final states.  The dotted portions of the curves 
    are potentially affected by solar modulation, and are thus subject to sizable systematic uncertainties. From Ref.~\cite{Bergstrom:2013jra}.}
  \label{fig:fraction}
\end{figure}

Even if dark matter is not responsible for the excess positrons observed by PAMELA and AMS-02, it is possible to use these measurements to place constraints on annihilating dark matter, in particular in the case of annihilations to charged leptons. In Fig.~\ref{fig:fraction} we show the constraints that result from the lack of a distinctive feature in the cosmic-ray positron spectrum~\cite{Bergstrom:2013jra,Ibarra:2013zia}. For dark matter that annihilates to $e^+ e^-$ ($\mu^+ \mu^-$), this constraint rules out the thermal relic benchmark cross section for masses up to $\sim$\,170 GeV ($\sim$\,100 GeV).

\subsection{Cosmic-Ray Antiprotons}

\begin{figure}[t!]
  \includegraphics[width=0.49\linewidth]{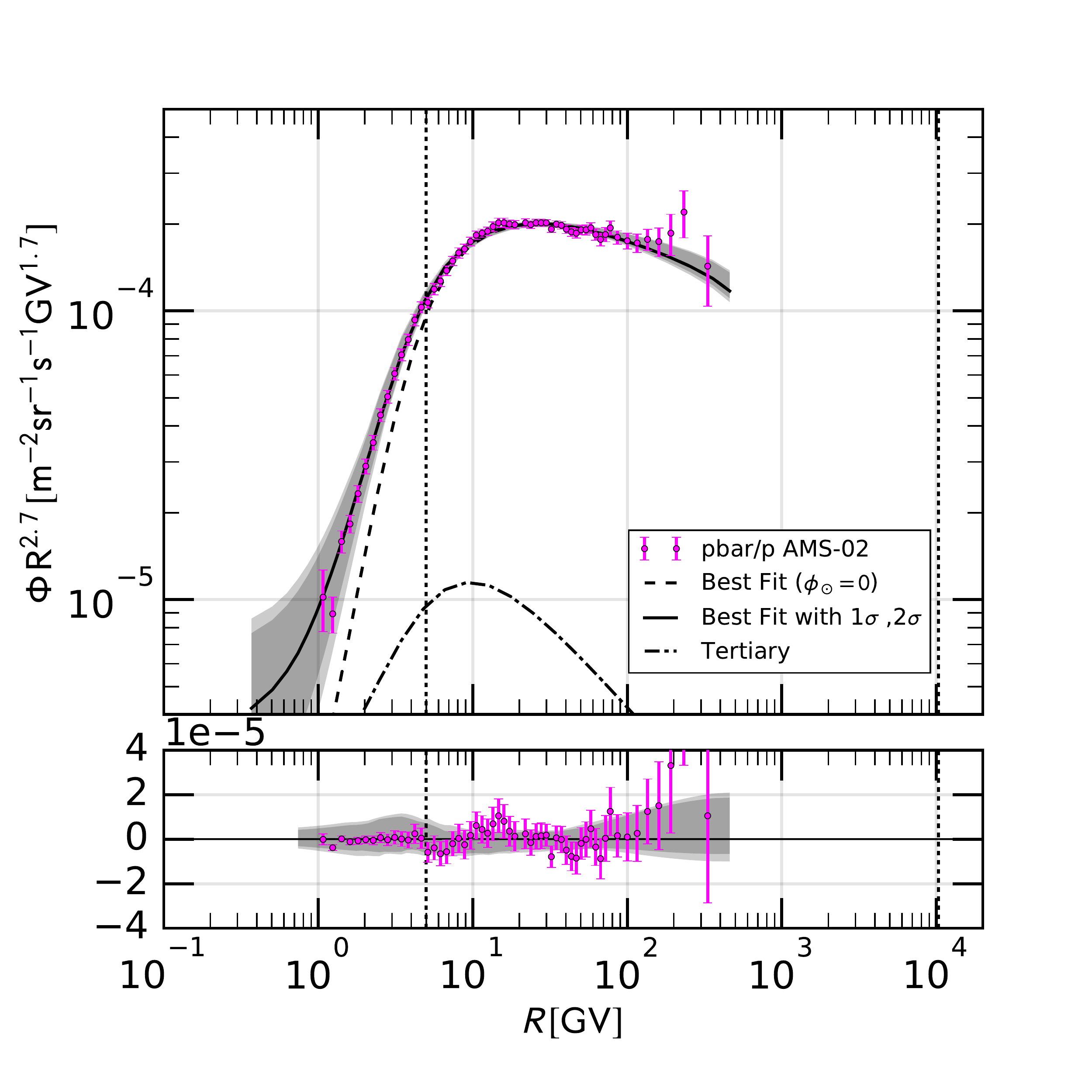}
  \includegraphics[width=0.49\linewidth]{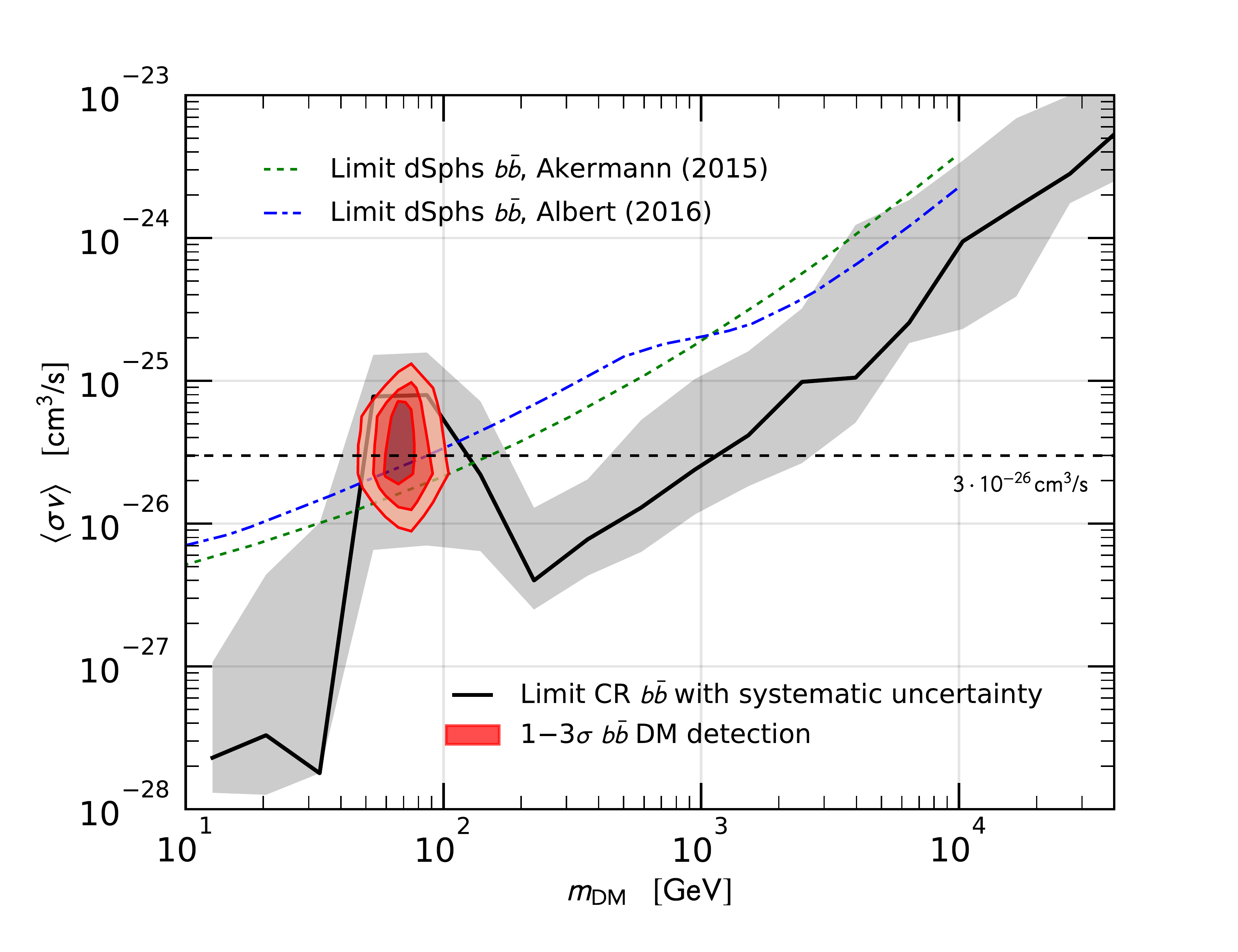}
  \caption{Left: The cosmic-ray antiproton-to-proton ratio as a function of rigidity as measured by AMS-02 compared to that predicted from standard secondary production in the interstellar medium. The lower panel shows the corresponding residual, with the grey bands representing the 1 and 2$\sigma$ uncertainties. Although an excess appears at energies between 10 and 20 GeV, systematic uncertainties associated with the antiproton production cross section, solar modulation and cosmic-ray transport make the significance of such features difficult to assess. Right: Constraints on the dark matter annihilation cross section (for annihilations to $b\bar{b}$) from the $\bar{p}/p$ ratio. In this frame the grey bands represent the range of constraints that are derived for various assumptions, and can be treated as an estimate of the systematic uncertainties. From Ref.~\cite{Cuoco:2016eej}.}
\label{fig:antiprotonlimits}
\end{figure}

In addition to positrons, the AMS-02 experiment has also produced a high-precision measurement of the cosmic-ray antiproton spectrum~\cite{Aguilar:2016kjl}. Analysis of the antiproton-to-proton ratio, in conjunction with other secondary-to-primary ratios, has found overall agreement with the expectations for standard secondary production over much of the measured energy range. At energies between 10 and 20 GeV and above $100$ GeV, however, there appears to be an excess of antiprotons (see Fig.~\ref{fig:antiprotonlimits})~\cite{Cuoco:2016eej,Cui:2016ppb} (see also Refs.~\cite{Cui:2018klo,Cuoco:2017rxb,Cuoco:2017iax}). At the highest energies, this excess could quite plausibly be the result of the reacceleration of antiproton secondaries produced in supernova remnants~\cite{Cholis:2017qlb}. The excess at 10-20 GeV has no simple explanation, however, and has been interpreted as a possible signal of annihilating dark matter~\cite{Cuoco:2016eej,Cui:2016ppb,Cui:2018klo,Cuoco:2017rxb,Cuoco:2017iax}. That being said, systematic uncertainties related to the antiproton production cross section, solar modulation and cosmic-ray transport make the significance of this feature difficult to assess at this time~\cite{Reinert:2017aga,Winkler:2017xor}. Even so, it is intriguing to note that the range of dark matter models favored by the Galactic Center gamma-ray excess are also well suited to produce an antiproton excess similar to that measured by AMS-02 ($m_X\sim 60-80$ GeV, $\sigma v \sim 2 \times 10^{-26}$ cm$^3/$s). In Fig.~\ref{fig:antiprotonlimits} we show the antiproton-to-proton ratio as measured by AMS-02 and the resulting constraints on annihilating dark matter (as well as the region of parameter space favored to produce the observed excess).

\subsection{Anti-Deuterium and Anti-Helium}

Although AMS-02 has not yet published the results of their searches for anti-deuterium or anti-helium events (see, however, Ref.~\cite{AMSLaPalma}), these channels could potentially provide a powerful probe of annihilating dark matter~\cite{Donato:1999gy,Carlson:2014ssa,Cirelli:2014qia,Coogan:2017pwt,Korsmeier:2017xzj}. Given the very low fluxes of anti-deuterium and anti-helium that are predicted from astrophysical sources or mechanisms, even a handful of such events could constitute a strong signal of annihilating dark matter (or other new physics). It has even been argued that the observation of a single cosmic-ray anti-deuteron with a rigidity below 1 GV would constitute a compelling signal of annihilating dark matter~\cite{Donato:1999gy,Fuke:2005it,Donato:2008yx,Ibarra:2013qt,Hryczuk:2014hpa,Carlson:2014ssa,Aramaki:2015laa,Reinert:2017aga}.

There exist, however, very substantial uncertainties related to the anti-nuclei fluxes predicted from standard secondary production, as well as from annihilating dark matter. Although these uncertainties make the prospects for such searches somewhat difficult to assess, measurements of cosmic-ray nuclei by AMS-02 (as well as GAPS~\cite{Aramaki:2015laa,Ong:2017szd}) are generally expected to be among the most exciting channels for indirect dark matter searches in the years ahead.

\section{Neutrino Searches for Dark Matter Annihilation Products}

In addition to gamma rays and cosmic rays, dark matter annihilations can generate high-energy neutrinos, potentially detectable by telescopes such as IceCube~\cite{Aartsen:2016zhm,Aartsen:2016fep} or, at lower energies, Super-Kamiokande~\cite{Desai:2004pq}. Strategies similar to those described for gamma-ray telescopes in Sec.~\ref{gammasec} have been employed to use neutrino telescopes to search for dark matter annihilation products from the Galactic Center~\cite{Aartsen:2017ulx}, or the Galactic Halo ~\cite{Aartsen:2016pfc}. Due to the small interaction cross sections of neutrinos, however, such constraints are in most cases much weaker than those derived from gamma-ray or cosmic-ray based searches, generally leading to upper limits on the annihilation cross section that lie between $\langle \sigma v \rangle \sim 10^{-21} - 10^{-23}$ cm$^3/$s, depending on the mass of the dark matter candidate and the annihilation channel. 

Neutrinos do, however, have a potential advantage over gamma rays and cosmic rays in that they can penetrate large quantities of matter. As a result, it may be possible to detect neutrinos that are produced through dark matter annihilations in the core of the Sun or Earth~\cite{Silk:1985ax,Hagelin:1986gv,Freese:1985qw,Krauss:1985aaa,Gaisser:1986ha}. Unlike most other indirect searches, which depend primarily on the dark matter's annihilation cross section, the prospects for detecting such annihilations in the core of the Sun or Earth also depend in large part on the dark matter's capture rate, and thus on its elastic scattering cross section with nuclei. Although the full calculation of the capture rate is involved~\cite{Gould:1987ir}, we can make a simple back-of-the-envelope estimate for the solar capture rate as follows:
\begin{eqnarray}
C^{\odot} &\sim& \phi_X (M_{\odot}/m_p) \sigma_{Xp}, \\
&\sim& 10^{20} \, {\rm s}^{-1} \times \bigg(\frac{100 \, {\rm GeV}}{m_X}\bigg) \bigg(\frac{\sigma_{Xp}}{10^{-42} \, {\rm cm}^2}\bigg), \nonumber
\end{eqnarray}
where $\phi_X$ is the flux of dark matter particles in the Solar System, $M_{\odot}$ is the mass of the Sun, and $\sigma_{Xp}$ is the dark matter-proton elastic scattering cross section. In the lower line of this expression, we have adopted reasonable values for the local density ($\rho_X = 0.3$ GeV/cm$^3$) and velocity distribution ($\bar{v}=270$ km/s) of dark matter particles. A more careful calculation, including the effects of gravitational focusing and the probability that a scattered dark matter particle will ultimately be gravitationally bound, leads to the following solar capture rate~\cite{Gould:1987ir}:
\begin{eqnarray}
C^{\odot} &\approx& 1.3\times 10^{21} \, {\rm s}^{-1} \times \bigg(\frac{100 \, {\rm GeV}}{m_X}\bigg) \sum_i \bigg(\frac{A_i \, \sigma_{Xp} \,S(m_X/m_i)}{10^{-42} \, {\rm cm}^2}\bigg),
\end{eqnarray}
where $A_i$ denotes the relative abundance of each nuclear species, $A_{\rm H}=1.0$, $A_{\rm He}=0.07$, $A_{\rm O} = 0.0005$, etc. The quantity $S(m_X/m_i)$ is a kinematic factor, defined as follows:
\begin{equation}
S(x) = \bigg[\frac{A(x)^{3/2}}{1+A(x)^{3/2}}\bigg]^{2/3},
\end{equation}
where
\begin{equation}
A(x) =\frac{3}{2} \frac{x}{(x-1)^2} \bigg(\frac{v_{\rm esc}}{\bar{v}}\bigg)^2,
\end{equation}
and $v_{\rm esc} \simeq 1156$ km/s is the escape velocity of the Sun. Notice that for dark matter particles much heavier than their nuclear targets, $S \propto 1/m_X$, kinematically suppressing the overall capture rate. 

The number of dark matter particles present in the Sun as a function of time is given as follows:
\begin{equation}
\dot{N}(t) = C^{\odot} -A^{\odot} N(t)^2 -E^{\odot} N,
\label{diffeq}
\end{equation}
where $C^{\odot}$ is the capture rate described above, $A^{\odot}$ is the dark matter's annihilation cross section, $\langle \sigma v \rangle$, divided by the effective volume that is occupied by the captured dark matter, and $E^{\odot}$ is inverse time for a dark matter particle to escape the Sun by evaporation. The effective volume is determined by matching the temperature of the Sun's core to the gravitational potential energy of a single dark matter particle:
\begin{equation}
V_{\rm eff} \simeq 5.7 \times 10^{27} \, {\rm cm}^3 \, \bigg(\frac{100 \, {\rm GeV}}{m_X}\bigg)^{3/2}.
 \end{equation}

For dark matter particles heavier than a few GeV, evaporation is negligible. In this case, the solution to Eq.~\ref{diffeq} can be written as follows:
\begin{equation}
\Gamma(t) = \frac{1}{2}A^{\odot}N(t)^2 = \frac{1}{2}C^{\odot} \tanh^2(t \sqrt{C^{\odot} A^{\odot}}),
\end{equation}
where $\Gamma(t)$ is the present annihilation rate of dark matter particles as a function of the age of the Sun. Notice that for $t \gg (C^{\odot} A^{\odot})^{-1/2}$ the annihilation rate becomes a constant, $\Gamma = C^{\odot}/2$, having reached an equilibrium between the rates of capture and annihilation.

\begin{figure}[t!]
  \includegraphics[width=0.7\linewidth]{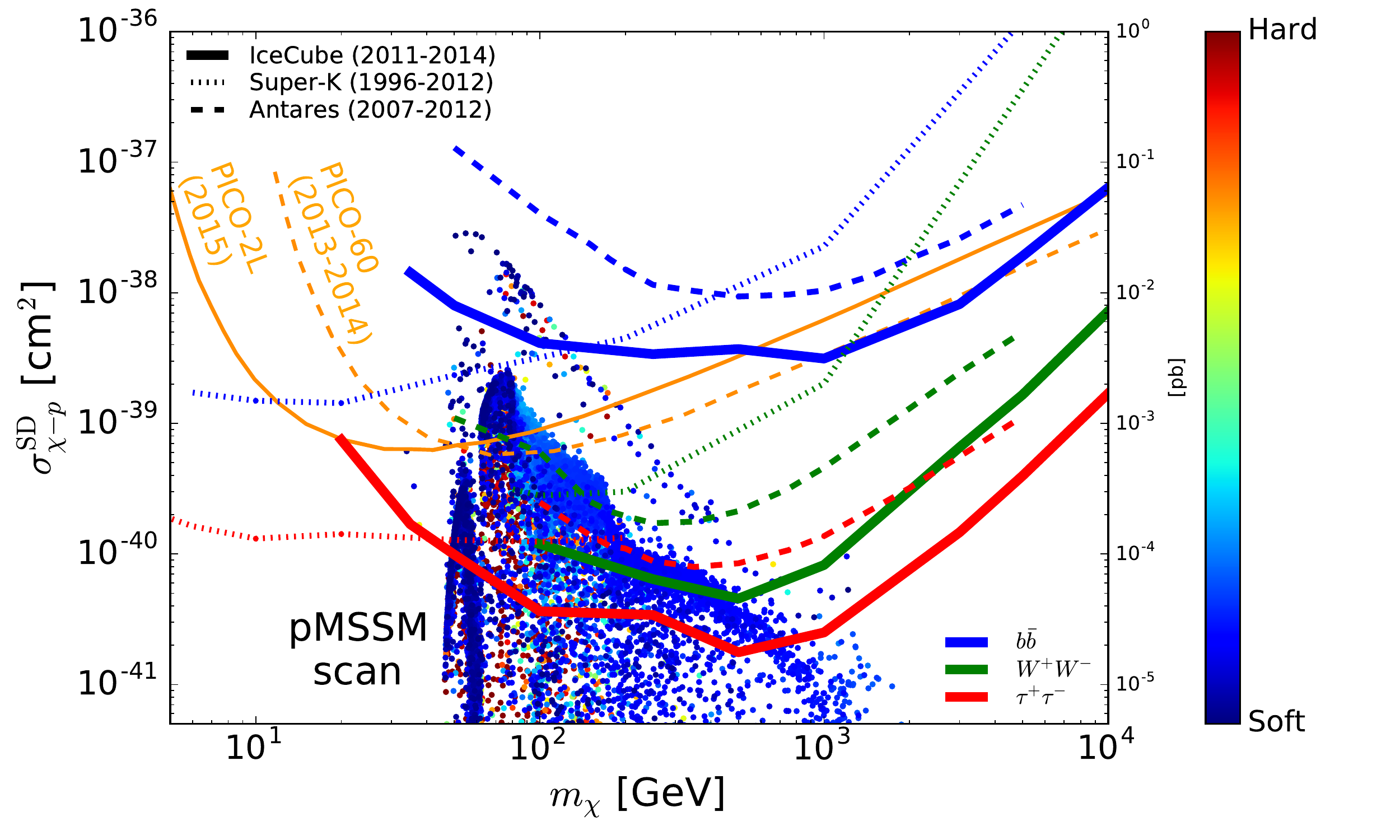}
  \caption{Constraints on the dark matter's spin-dependent elastic scattering cross section with protons as derived from observations of the Sun by IceCube, Antares and Super-Kamiokande. From Ref.~\cite{Aartsen:2016zhm}.}
\label{fig:IceCube}
\end{figure}

In Fig.~\ref{fig:IceCube}, we show constraints from the IceCube, Antares and Super-Kamiokande experiments on dark matter particles annihilating in the core of the Sun~\cite{Aartsen:2016zhm}. Notice that these constraints are not on the dark matter's annihilation cross section, but on its elastic scattering cross section with nuclei. In fact, these constraints are derived under the assumption of capture-annihilation equilibrium, $\Gamma = C^{\odot}/2$, which one should expect to be realized for the range of elastic scattering cross sections shown so long as $\langle \sigma v \rangle \gsim 10^{-27}$ cm$^3/$s. Furthermore, the constraints shown are for the case of spin-dependent scattering with nuclei, as the constraints from direct detection experiments on spin-independent scattering are very stringent, and rule out most models that neutrino telescopes would be sensitive to. From this figure, we see that the constraints from neutrino telescopes can exceed those from direct detection experiments in cases in which the dark matter annihilates to final states that produce large numbers of high-energy neutrinos, such as $W^+ W^-$ or $\tau^+ \tau^-$.

\section{Constraints on Annihilating Dark Matter from the Cosmic Microwave Background}

Thus far in these lectures, I have focused on ways in which we could potentially observed the annihilation products of dark matter directly. But it is also possible to place constraints on dark matter by studying the impact of their annihilation products on the universe during various eras of cosmic history. In particular, dark matter annihilation products could alter the light element abundances that are produced during Big Bang Nucleosynthesis (BBN), or change the ionization history of our universe during and after the formation of the cosmic microwave background (CMB).

Consider a thermal relic dark matter candidate. By the definition of what it means for a particle to undergo freeze-out, an order one fraction of the total dark matter population underwent annihilations in a Hubble time during this process. As the universe continued to expand, the annihilation rate dropped rapidly. We can write the number of dark matter annihilations per comoving volume per Hubble time as follows:
\begin{equation}
N_{\rm ann} = \frac{1}{2}\frac{\rho^2_X \langle \sigma v \rangle V_c}{m^2_X H},
\end{equation}
where $V_c$ is the moving volume and $H$ is expansion rate (making $1/H$ the Hubble time). Since $\rho_X \propto a^{-3}$, $V_c \propto a^3$, and $H \propto g^{1/2}_{\star} a^{-2}$ (during radiation domination), we conclude that the fraction of annihilations per Hubble time evolved as $N_{\rm ann} \propto g^{-1/2}_{\star} a^{-1}$ up until matter-radiation equality, and as $N_{\rm ann} \propto g^{-1/2}_{\star} a^{-3/2}$ during matter domination.  

For concreteness, consider a dark matter candidate with a mass of 100 GeV and that froze-out when the temperature was $T_{\rm FO} \simeq 100\, {\rm GeV} /20 \simeq 5$ GeV. From the scaling relationship described above, we estimate that by the time that the universe has cooled to a temperature of 1 eV, on the order of 0.1 eV per baryon was injected into the universe through dark matter annihilations per Hubble time. This is enough energy to ionize up to $\sim (0.1 \,{\rm eV}) /(13.7 \,{\rm eV}) \sim 10^{-3}$ of the hydrogen atoms, substantially impacting the process of recombination, and well as the observed characteristics of the CMB. In fact, measurements of the CMB allow us to place stringent and robust constraints on thermal relic dark matter candidates, excluding those with velocity independent annihilation cross sections with masses up to $\sim$\,10-30 GeV~\cite{Slatyer:2015jla,Galli:2013dna,Finkbeiner:2011dx,Galli:2011rz,Galli:2009zc,Slatyer:2009yq}.

\begin{figure}[t!]
  \includegraphics[width=1.0\linewidth]{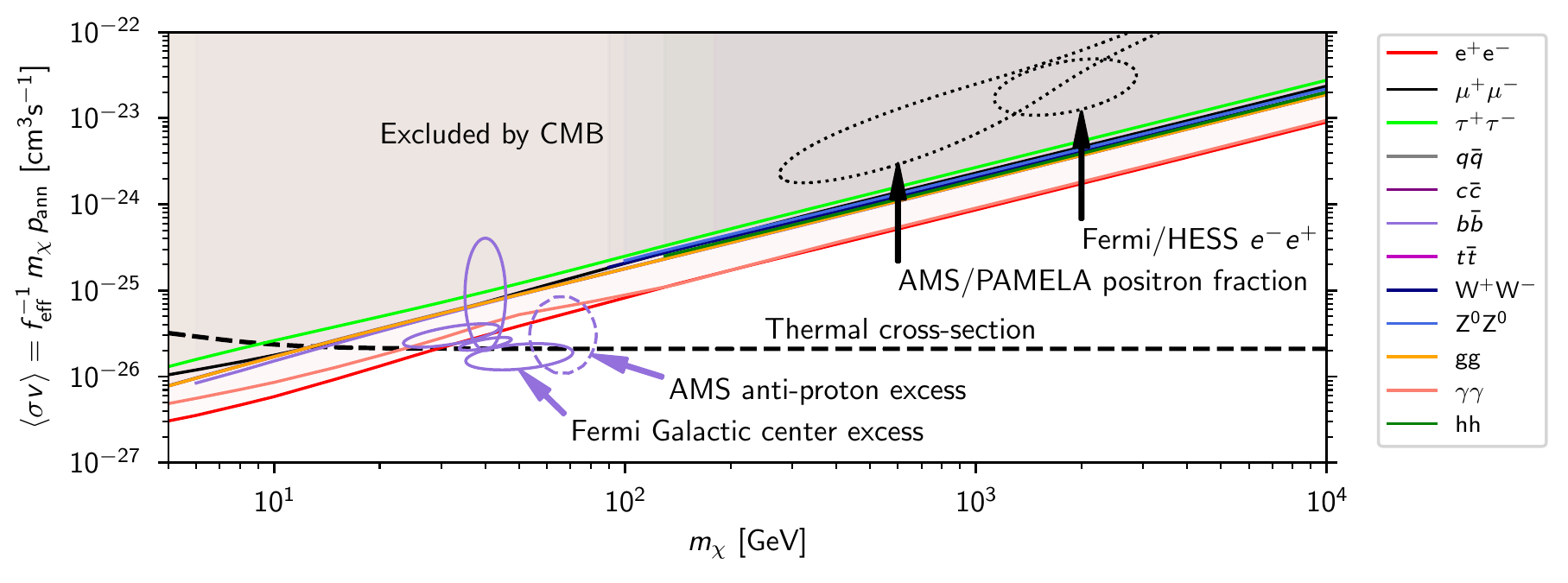}
  \caption{Constraints on the dark matter annihilation cross section (for a variety of annihilation channels) from the Planck Collaboration's measurements of the cosmic microwave background. From Ref.~\cite{Aghanim:2018eyx}.}
\label{fig:planck}
\end{figure}

 In Fig.~\ref{fig:planck}, we show the most recent constraints on annihilating dark matter from the Planck Colaboration~\cite{Aghanim:2018eyx}. Although these constraints do not extend to masses as high as some of the others discussed in these lectures, they are very robust and suffer from negligible astrophysical or systematic uncertainties. Furthermore, whereas gamma-ray and cosmic-ray searches for dark matter are generally less sensitive at masses below $\sim$\,10 GeV or so, CMB constraints rely only on the total electromagnetic power injected and thus extend to masses well below the range shown in Fig.~\ref{fig:planck}. The CMB-based constraints are strongest for dark matter candidates which annihilate to electrons or photons, as these channels deposit the largest quantities directly into heating and ionizing the intergalactic medium. Lastly, this figure also identifies regions of parameter space in which dark matter could account for the Galactic Center gamma-ray excess, the cosmic-ray antiproton excess, or cosmic-ray positron excess, as discussed earlier in these lectures.

\section{Decaying Dark Matter}

So far, these lectures have focused on searches for dark matter annihilation products. This choice was motivated in part by the arguments presented in Sec.~\ref{sectionone}, which relate the abundance of dark matter to the annihilation cross section of a thermal relic. But despite these arguments, there are many examples of viable dark matter candidates which do not appreciably annihilate. Alternatively, the particles that make up the dark matter could be unstable, and produce potentially observable fluxes of decay products. 

Observations of the the cosmic microwave background (CMB) and large scale structure indicate that the abundance of dark matter has not appreciably changed over the course of the matter-dominated era of our universe's history. In fact, even if the decay products of dark matter are invisible, such measurements can be used to constrain $\tau_{X} \gsim 2 \times 10^{19}$ s~\cite{Poulin:2016nat}. Much stronger constraints can be placed on dark matter candidates that decay into detectable particles.

Unlike in the case of dark matter annihilation, there is no clear benchmark target for the lifetime of a long-lived but unstable dark matter particle. That being, arguments have been made which favor some ranges of lifetimes. For example, the lifetime of a particle that decays through a dimension-5 operator suppressed by the GUT scale ($M_{\rm GUT} \sim 10^{16}$ GeV) can be estimated as follows:
\begin{equation}
\tau \sim \frac{M^2_{\rm GUT}}{m^3_{X}} \sim 10^{17} \, {\rm s} \times \bigg(\frac{{\rm MeV}}{m_{\rm DM}}\bigg)^3.
\end{equation}
From this we learn that dimension-5 operators, even if suppressed by a very high-scale, tend to cause dark matter particles to decay on timescales that are already ruled out by cosmological considerations, unless very light. On the other hand, if we consider a decay that results from a dimension-6 operator, we arrive at the following estimate:
\begin{equation}
\tau \sim \frac{M^4_{\rm GUT}}{m^5_{X}} \sim 10^{25} \, {\rm s} \times \bigg(\frac{{\rm TeV}}{m_{\rm DM}}\bigg)^3.
\end{equation}
This lifetime is not excluded on cosmological grounds, but could potentially be tested through searches for the dark matter's decay products (for a review, see Ref.~\cite{Ibarra:2013cra}). Searches for dark matter decay products in the form of gamma-rays~\cite{Blanco:2018esa,Cohen:2016uyg,Ando:2015qda,Hutsi:2010ai,Murase:2015gea,Murase:2012xs,Ackermann:2015lka,Ackermann:2012rg,Kalashev:2016cre,Cirelli:2012ut,Esmaili:2015xpa,Liu:2016ngs}, X-rays~\cite{Boyarsky:2007ge,Yuksel:2007xh,Perez:2016tcq}, neutrinos~\cite{Murase:2012xs,PalomaresRuiz:2007ry} and cosmic rays~\cite{Ibarra:2013zia} have each been carried out. 

To calculate the flux of gamma-rays from decaying dark matter, we modify Eq.~\ref{gamma}, replacing the annihilation rate per volume ($ \langle \sigma v \rangle \rho_X/2m_X^2$) with the decay rate per volume ($\rho_X/m_X \tau_X$), and substituting the gamma-ray spectrum produced per annihilation with that produced per decay:
\begin{equation}
\frac{dN_{\gamma}}{dE_{\gamma}} (E_{\gamma}, \Delta \Omega) = \bigg(\frac{dN_{\gamma}}{dE_{\gamma}}\bigg)  \frac{1}{4\pi \tau_{X} m_{X}} \int_{\Delta \Omega} \int_{los} \rho_{X}(l,\Omega) dl d\Omega.
\label{los}
\end{equation}
Because this flux is proportional to only one power of the dark matter density (as opposed to two in the case of dark matter annihilation), the best strategy is generally to study large regions of the sky, searching for decay products from throughout the halo of the Milky Way, and throughout the integrated volume of the observable universe. Due to the universe's opacity to gamma rays above $\sim$\,1 TeV, the constraints from Fermi on the dark matter's lifetime are approximately flat from the GeV scale to EeV masses and above, excluding decays to (non-neutrino) Standard Model particles for $\tau \lsim 10^{28}$ s.

\subsection{X-Ray Lines from Decaying Sterile Neutrinos}

\begin{table}
	\begin{center}
	\begin{tabular}{cc}
	\begin{tikzpicture}
		  \begin{feynman}
			\vertex (a1) {\(\nu_S\)};
			\vertex[right=1cm of a1] (a2);
			\vertex[above right=1cm and 2cm of a2] (c1);
			\vertex[below right=1cm and 2cm of a2] (b1);
			\vertex[right=1cm of c1] (c2)  {\(\gamma\)};
			\vertex[right=1cm of b1] (b2) {\(\nu\)};
			\diagram*{
				(a2) -- [fermion,  edge label'=\(l^-\)] (c1);
				(c1) -- [fermion,  edge label'=\(l^-\)] (b1);
				(a2) -- [boson, edge label'=\(W^+\)] (b1);	
				(a1) -- [fermion] (a2);
				(b1) -- [fermion] (b2);
				(c1) -- [boson] (c2);		
			};
		  \end{feynman}
		\end{tikzpicture} & 
		\begin{tikzpicture}
		  \begin{feynman}
			\vertex (a1) {\(\nu_S\)};
			\vertex[right=1cm of a1] (a2);
			\vertex[above right=1cm and 2cm of a2] (c1);
			\vertex[below right=1cm and 2cm of a2] (b1);
			\vertex[right=1cm of c1] (c2)  {\(\gamma\)};
			\vertex[right=1cm of b1] (b2) {\(\nu\)};
			\diagram*{
				(a2) -- [boson,  edge label'=\(W^+\)] (c1);
				(b1) -- [boson,  edge label'=\(W^+\)] (c1);
				(a2) -- [fermion, edge label'=\(l^-\)] (b1);	
				(a1) -- [fermion] (a2);
				(b1) -- [fermion] (b2);
				(c1) -- [boson] (c2);		
			};
		  \end{feynman}
		\end{tikzpicture}\\
	\end{tabular}
	\end{center}
	
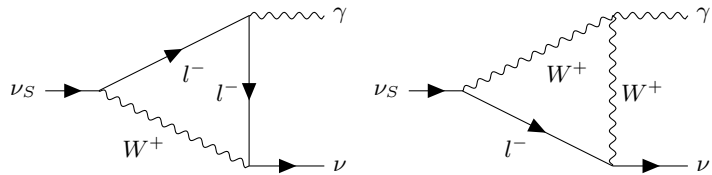
\captionof{figure}{Feynman diagrams for the decay of a sterile neutrino to a photon and a neutrino.}
	\label{fig:neutrino}
\end{table}

The origin of neutrino masses remains one of the most important outstanding puzzles in particle physics. Although the Standard Model does not accommodate masses for these particles, natural extensions can easily generate small masses for these species through variations of the see-saw mechanism~\cite{Schechter:1980gr,GellMann:1980vs,Mohapatra:1979ia,Yanagida:1979as,Minkowski:1977sc}. Such scenarios predict the existence of sterile neutrinos, which do not interact through the weak force. If the degree of mixing between the sterile and active neutrinos is very small, the sterile neutrinos will not reach thermal equilibrium with the Standard Model bath in the early universe. As pointed out by Dodelson and Widrow~\cite{Dodelson:1993je}, however, even a very small degree of mixing can generate a significant population of sterile neutrinos through the collisions of active neutrinos with other Standard Model particles (see also Refs.~\cite{Barbieri:1989ti,Kainulainen:1990ds}). Sterile neutrinos with masses in the range of $\sim$\,1-100 keV have long been considered as potentially viable candidates for dark matter (for a review, see Ref.~\cite{Adhikari:2016bei}).

In recent years, this framework has become increasingly constrained. In particular, sterile neutrinos can decay to a final state that includes a distinctive mono-energetic photon through diagrams of the kind shown in Fig.~\ref{fig:neutrino}. With this signal in mind, searches for X-ray~\cite{Boyarsky:2007ge,Yuksel:2007xh,Perez:2016tcq} and gamma-ray~\cite{Ackermann:2015lka} lines have resulted in strong upper limits on the lifetime of sterile neutrinos, which in turn constrains the mixing angle between the sterile and active species. When these results are combined with observations associated with structure formation~\cite{Horiuchi:2013noa, Schneider:2016uqi}, one finds that sterile neutrinos within the context of the standard Dodelson-Widrow scenario are unable to account for the entirety of the cosmological dark matter abundance. 

In light of these constraints, a number of less minimal scenarios have been proposed in which the production rate of sterile neutrinos is enhanced in the early universe, allowing for smaller mixing angles and thus relaxing the constraints from astrophysical observations. Model-building efforts in this direction have generally relied on either resonant enhancements or additional out-of-equilibrium processes. The former can be realized with the inclusion of a non-negligible lepton asymmetry in the early universe, which effectively modifies the matter potential of the Standard Model neutrinos~\cite{Shi:1998km}. In this case, the successful predictions of Big Bang nucleosynthesis limit the degree to which the mixing can be suppressed, and only a small window of parameter space remains phenomenologically viable, corresponding to sterile neutrinos in the mass range of approximately 7--25 keV~\cite{Perez:2016tcq}. Alternatively, the second class of models explicitly incorporates new particle species, such as additional scalars that decay directly into dark matter~\cite{Kusenko:2006rh,Merle:2013wta,Frigerio:2014ifa,Merle:2015oja,Shaposhnikov:2006xi,Petraki:2007gq,Adulpravitchai:2014xna,Frigerio:2014ifa,Kadota:2007mv,Abada:2014zra,Shuve:2014doa}. In these scenarios, the connection between the production and late-time decays of sterile neutrinos is blurred, essentially at the cost of introducing additional degrees-of-freedom that are not directly tied to sterile-active oscillations. 

Interest in sterile neutrino dark matter has been bolstered in recent years by the reported detection of a 3.5 keV X-ray line from a stacked collection of galaxy clusters using data from XMM-Newton~\cite{Bulbul:2014sua,Boyarsky:2014jta}. More recently, the presence of a similar line has been detected from the center of the Milky Way~\cite{Boyarsky:2014ska} and in deep-field observations~\cite{Cappelluti:2017ywp}. The analysis of X-ray data from the direction of the Draco dwarf galaxy as described in Ref.~\cite{Jeltema:2015mee} appears to rule out the presence of such a signal, while the authors of Ref.~\cite{Ruchayskiy:2015onc} claim to have detected a faint 3.5 keV line signal in the same dataset. The lack of such a line feature in the emission from the Andromeda Galaxy~\cite{Horiuchi:2013noa}, a stacked sample of galaxies~\cite{Anderson:2014tza}, and dwarf galaxies~\cite{Malyshev:2014xqa} has been used to establish strong limits on dark matter related interpretations of this signal. While some groups have argued that spectral lines from hot potassium or chlorine gas in the intercluster medium might be responsible for this signal, this interpretation remains actively debated~\cite{Jeltema:2014qfa, Boyarsky:2014paa, Bulbul:2014ala, Jeltema:2014mla}. For a review, see Ref.~\cite{Abazajian:2017tcc}.

Although a 7 keV decaying sterile neutrino is among the most well-motivated explanations for the observed 3.5 keV line, the constraints mentioned in the above paragraph have cast some doubt on this interpretation. With this in mind, a number of alternatives have been proposed. For example, scenarios have been considered in which pairs of dark matter particles can scatter to excite one (or both) into a slightly heavier state, which then produces a 3.5 keV photon in its subsequent decay into the ground state~\cite{Finkbeiner:2014sja,Cline:2014vsa}. But whereas a decaying sterile neutrino would produce a 3.5 keV signal in proportion to its density, such an ``exciting dark matter'' scenario leads to a signal that scales with the square of the density, along with some dependence on the dark matter's velocity distribution. This could potentially provide an explanation for why no 3.5 keV signal has been observed from dwarf galaxies.

\section{Summary}

In these lectures, I have presented a overview of indirect searches for dark matter, describing searches for gamma rays, cosmic rays and neutrinos from dark matter annihilations or decays, and the impact of such particles on the cosmic microwave background. It should be noted that these lectures are far from exhaustive, and there are many efforts to detect dark matter indirectly that I have not discussed here. A few takeaways from these lectures are the following:
\begin{itemize}
\item{The measured abundance of dark matter provides us with motivation to consider dark matter candidates that annihilate with a cross section near the benchmark value of $\langle \sigma v \rangle \simeq 2 \times 10^{-26}$ cm$^3/$s. Furthermore, we can restrict the mass of thermal relics to be heavier than a few MeV (in order to satisfy constraints from Big Bang Nucleosynthesis) and lighter than 120 TeV (in order to not violate partial wave unitarity).}
\item{Measurements of the cosmic microwave background have been used to place constraints on annihilating dark matter, excluding most candidates with the thermal relic benchmark cross section for masses up to $m_X \simeq 10-30$ GeV.}
\item{Gamma-ray observations of dwarf galaxies and the Galactic Center extend these limits up to $m_X \sim 60-140$ GeV. If taken at face value, the cosmic-ray antiproton spectrum appears to exclude such candidates for masses between $m_X \sim160-500$ GeV, although significant systematic uncertainties apply to this channel. For the case of dark matter annihilations to $e^+ e^-$, the cosmic-ray positron spectrum also provides strong constraints.} 
\item{A number of excesses and anomalies have been reported which could be the result of dark matter annihilations or decays. In particular, the Galactic Center gamma-ray excess and the cosmic-ray antiproton excess each point toward dark matter annihilating with a cross section near the benchmark value of $\langle \sigma v \rangle \simeq 2 \times 10^{-26}$ cm$^3/$s and with a mass in the approximate range of $\sim$\,50-80 GeV (for the representative example of annihilations to $b\bar{b}$).}
\item{There are many viable models in which the dark matter annihilates with a cross section that is significantly smaller that the thermal relic benchmark (as a consequence of $p$-wave annihilations, coannihilations, non-standard cosmological histories, etc.). At present, indirect searches for dark matter are not generally sensitive to such scenarios.}
\end{itemize}

Looking forward, we expect indirect searches for dark matter to be bolstered by a range of new experiments and observations. The CTA is an array of ground-based gamma-ray telescopes scheduled for construction between 2020 and 2025, offering unprecedented sensitivity to the very high-energy gamma-ray sky~\cite{Carr:2015hta,Doro:2012xx}. At lower gamma-ray energies are the proposed satellite-based AMIGO and e-ASTROGAM telescopes, which are designed to be significantly more sensitive than Fermi at energies below 1 GeV~\cite{Chou:2017wrw,DeAngelis:2017gra,Bartels:2017dpb}. Searches for gamma rays from dark matter annihilation in dwarf galaxies will be further enhanced by LSST, which is expected to discover many new dwarfs. There is much yet to be learned about dark matter from the measurements of AMS-02, in particular in regards to their search for anti-deuterium and anti-Helium in the cosmic-ray spectrum. I expect this to be a exciting topic in the years ahead. Lastly, I will also mention that plans are underway to launch a satellite-based X-ray telescope with the spectral resolution required to strongly constrain the origin of the 3.5 keV line~\cite{Speckhard:2015eva}  (a replacement for the Hitomi satellite, which was lost in 2016).

\section*{Acknowledgements}
These lectures were originally presented at TASI 2018: ``Theory in an Era of Data'', which was supported by the U.S.~National Science Foundation. I would like to thank the organizers of TASI 2018 -- Tracy Slatyer, Tilman Plehn and Tom DeGrand -- as well as the students that participated. These lectures have been authored by Fermi Research Alliance, LLC under Contract No. DE-AC02-07CH11359 with the U.S. Department of Energy, Office of Science, Office of High Energy Physics. The United States Government retains and the publisher, by accepting the article for publication, acknowledges that the United States Government retains a non-exclusive, paid-up, irrevocable, world-wide license to publish or reproduce the published form of this manuscript, or allow others to do so, for United States Government purposes.

\bibliography{tasi.bib}
\end{document}